\documentstyle[12pt,thmsa,sw20lart]{article}

\input tcilatex
\begin{document}

\title{Particle Aspects of Cosmology \\
and \\
Baryogenesis\thanks{%
Lectures delivered at Department of Physics, KFUPM, Dhahran, Saudi Arabia
(April 2002 and September 2002).}}
\author{Riazuddin \\
National Center for Physics\\
Quaid-e-Azam University\\
Islamabad 45320 Pakistan\\
and\\
Department of Physics\\
King Fahd University of Petroleum and Minerals\\
Dhahran 31261 Saudi Arabia}
\maketitle

\begin{abstract}
An introduction to particle aspects of cosmology with particular refrence to
primordial nucleosynthesis, dark matter and baryeogenesis is provided. In
particular, various scenarios--Grand Unified Theory baryogenesis,
electroweak baryogenesis and baryogenesis through leptogenesis are reviewed.
\end{abstract}

\newpage\ 

\section{Introduction}

I will first discuss Nucleosynthesis and show how it leads to two big
problems in cosmology: Dark Matter and Baryogenesis. Before I discuss
Primordial Nucleo-synthesis, I give you some background:

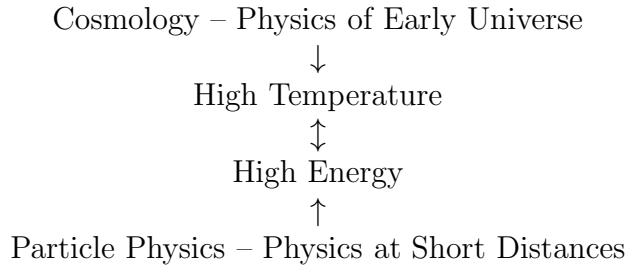
\begin{center}
Cosmology -- Physics of Early Universe

$\downarrow $

High Temperature

$\updownarrow $

High Energy

$\uparrow $

Particle Physics -- Physics at Short Distances
\end{center}

\section{Thermal Equilibrium}

Consider an arbitrary volume $V$ in thermal equilibrium with a heat bath at
temperature $T$. The particle density $n_{i}$ (i, particle index) at
temperature $T$ is given by 
\begin{equation}
n_{i}=\frac{N_{i}}{V}=\frac{g_{i}}{2\pi ^{2}}\left( \frac{k_{B}T}{\hbar c}%
\right) ^{3}\dint\nolimits_{0}^{\infty }\left[ \exp \left( \frac{E}{k_{B}T}%
\right) \pm 1\right] ^{-1}\ z^{2}dz.  \label{01}
\end{equation}
The energy density is given by 
\begin{equation}
\rho _{i}\ c^{2}=\frac{g_{i}}{2\pi ^{2}}\left( \frac{k_{B}T}{\hbar c}\right)
^{3}\left( k_{B}\ T\right) \dint\nolimits_{0}^{\infty }\left[ \exp \left( 
\frac{E}{k_{B}T}\right) \pm 1\right] ^{-1}\ \left( \frac{E}{k_{B}T}\right)
z^{2}dz^{2},  \label{02}
\end{equation}
where 
\begin{equation}
z=\frac{qc}{k_{B}T},\qquad E=\left[ \left( qc\right) ^{2}+\left(
m_{i}c^{2}\right) ^{2}\right] ^{1/2}
\end{equation}
and $g_{i}$ are the number of spin states, $q$ is the momentum of the
particle and $m_{i}$ is its mass. The + sign is for the fermions ($F$) and $%
- $ sign is for the bosons ($B$). In particular for $i=$ photon, $m=0,\ g=2$%
. In writing Eqs. (\ref{01}) and (\ref{02}), we have put the chemical
potential $\mu _{i}=0$. For photon $\mu =0$. Since particles and
antiparticles are in equilibrium with photons $\mu _{i}=-\mu _{\bar{i}}$ .
If there is no asymmetry between the number of particles and antiparticles, $%
\mu _{i}=\mu _{\bar{i}}=0$. If the difference between the number of
particles and antiparticles is small compared with the number of photons, 
\begin{equation}
\left| \frac{\mu _{i}}{k_{B}T}\right| =\left| \frac{\mu _{\bar{i}}}{k_{B}T}%
\right| \ll 1
\end{equation}
and the chemical potential can be neglected. For the photon gas, we get from
Eqs. (\ref{01}) and (\ref{02}) [$T_{0}=2.725^{0}$ K, subscript 0 denotes the
present value of the temperature of cosmic background (CMB) radiation] 
\begin{eqnarray}
n_{\gamma } &=&2\frac{\zeta \left( 3\right) }{\pi ^{2}}\left( \frac{k_{B}T}{%
\hbar c}\right) ^{3}=2\frac{1.2}{\pi ^{2}}\left( \frac{1}{\hbar c}\right)
^{3}\left( k_{B}T\right) ^{3}  \label{05} \\
n_{\gamma _{0}} &=&\frac{2.4}{\pi ^{2}}\left( \frac{1}{\hbar c}\right)
^{3}\left( k_{B}T_{0}\right) ^{3}=410.50\,\,\text{cm}^{-3} \\
n_{\gamma }\left( T\right) &=&410.50\left( \frac{T}{2.725}\right) ^{3}\,\,%
\text{cm}^{-3}, \\
\rho _{\gamma }c^{2} &=&6\frac{\zeta \left( 4\right) }{\pi ^{2}}\left( \frac{%
1}{\hbar c}\right) ^{3}\left( k_{B}T\right) ^{4}  \nonumber \\
&=&\frac{\pi ^{2}}{15}\left( \frac{1}{\hbar c}\right) ^{3}\left(
k_{B}T\right) ^{4}\simeq 2.7n_{\gamma }\left( k_{B}T\right) \\
\rho _{\gamma _{0}} &=&2.60\times 10^{-10}\,\,\text{GeV cm}^{-3}
\end{eqnarray}
The zeta functions are defined as follows 
\begin{eqnarray*}
\int_{0}^{\infty }\frac{z^{2}dz}{e^{z}-1} &=&\Gamma \left( 3\right) \zeta
\left( 3\right) \\
\int_{0}^{\infty }\frac{z^{2}dz}{e^{z}+1} &=&\left( 1-2^{-2}\right) \Gamma
\left( 3\right) \zeta \left( 3\right) =\frac{3}{4}\Gamma \left( 3\right)
\zeta \left( 3\right) \\
\int_{0}^{\infty }\frac{z^{3}dz}{e^{z}-1} &=&\Gamma \left( 4\right) \zeta
\left( 4\right) \\
\int_{0}^{\infty }\frac{z^{3}dz}{e^{z}+1} &=&\left( 1-2^{-3}\right) \Gamma
\left( 4\right) \zeta \left( 4\right) =\frac{7}{8}\Gamma \left( 4\right)
\zeta \left( 4\right)
\end{eqnarray*}
For a gas of extreme relativistic particles (ER), $k_{B}T\gg m_{i}c^{2}$, $%
qc\gg m_{i}c^{2}$, we thus get 
\begin{mathletters}
\begin{eqnarray}
n_{B} &=&\left( \frac{g_{B}}{2}\right) \ n_{\gamma },\qquad \rho _{B}=\left( 
\frac{g_{B}}{2}\right) \ \rho _{\gamma } \\
n_{F} &=&\frac{3}{4}\left( \frac{g_{F}}{2}\right) \ n_{\gamma },\qquad \rho
_{F}=\frac{7}{8}\left( \frac{g_{F}}{2}\right) \ \rho _{\gamma }.
\end{eqnarray}
The entropy $S$ for the photon gas is given by 
\end{mathletters}
\begin{equation}
S=\frac{R^{3}}{T}\frac{4}{3}\ \rho _{\gamma }\left( T\right) .
\end{equation}
For any relativistic gas 
\begin{equation}
S=\frac{R^{3}}{T}\frac{4}{3}\ \rho \left( T\right) .
\end{equation}
Thus for a gas consisting of extreme relativistic particles (bosons and
fermions): $(\hbar =c=1)$%
\begin{eqnarray}
n(T) &=&\frac{1}{2}\ g^{\prime }(T)\ n_{\gamma }(T)  \nonumber \\
&=&\frac{1.2}{\pi ^{2}}\ g^{\prime }(T)\ \left( k_{B}T\right) ^{3}
\label{13}
\end{eqnarray}
\begin{eqnarray}
\rho \left( T\right) &=&\frac{1}{2}\ g_{*}(T)\ \rho _{\gamma }(T)  \nonumber
\\
&=&\frac{\pi ^{2}}{30}\ g_{*}(T)\ \left( k_{B}T\right) ^{4}
\end{eqnarray}
\begin{equation}
S=\frac{R^{3}}{T}\frac{2}{3}\ g_{*}(T)\ \rho _{\gamma }(T),
\end{equation}
where 
\begin{mathletters}
\begin{eqnarray}
g^{\prime }(T) &=&\sum_{B}g_{B}+\frac{3}{4}\sum_{F}g_{F} \\
g_{*}(T) &=&\sum_{B}g_{B}+\frac{7}{8}\sum_{F}g_{F}
\end{eqnarray}
are called the ``effective'' degrees of freedom. We note that entropy per
unit volume is given by 
\end{mathletters}
\begin{equation}
\frac{1}{k_{B}}\frac{S}{R^{3}}=\frac{s}{k_{B}}=\frac{2\pi ^{2}}{45}\
g_{*}(T)\ \left( k_{B}T\right) ^{3}.
\end{equation}
For non-relativistic gas $k_{B}T\ll m_{i}c^{2}$, we use the Boltzmann
distribution 
\begin{equation}
n_{i}=\frac{g_{i}}{2\pi ^{2}}\left( \frac{k_{B}T}{\hbar c}\right) ^{3}\
\dint\nolimits_{0}^{\infty }\exp \left( -\frac{E}{k_{B}T}\right) z^{2}\ dz
\end{equation}
\begin{equation}
E\approx m_{i}c^{2}\ \left[ 1+\frac{1}{2}\frac{q^{2}c^{2}}{\left(
m_{i}c^{2}\right) ^{2}}\right] .
\end{equation}
This gives 
\begin{equation}
n_{i}=\left[ \frac{g_{i}}{\left( 2\pi \right) ^{3/2}}\right] \left( \frac{%
k_{B}T}{\hbar c}\right) ^{3}\left[ \left( \frac{m_{i}c^{2}}{k_{B}T}\right)
^{3/2}\ e^{-m_{i}c^{2}/k_{B}T}\right]  \label{20}
\end{equation}
\begin{equation}
\rho _{i}=n_{i}\ m_{i}.
\end{equation}

Expansion rate is given by the Hubble Parameter 
\[
H=\frac{\dot{R}}{R} 
\]
where [$R\left( t\right) $ is a scale factor for distances in co--moving
coordinates and describes the expansion of the universe] 
\begin{equation}
\dot{R}^{2}=\frac{8\pi G_{N}\rho }{3}R^{2}-kc^{2}+\frac{\Lambda c^{2}R^{2}}{3%
}  \label{22}
\end{equation}
The second term on the right hand side is due to curvature of space while
the third term containing the cosmological constant $\Lambda $, which being
very small ($\left| \Lambda \right| <3\times 10^{-52}$ m$^{-2}$) is usually
neglected. There is now evidence that $k=0$ (in any case for early universe $%
\rho \sim 1/R^{4}$ and as such the second and third terms on r.h.s. of Eq. (%
\ref{22}) can be neglected compared to the first term). Thus 
\begin{equation}
H=\frac{\dot{R}}{R}\simeq \left( \frac{8\pi G_{N}\rho }{3}\right) ^{1/2}
\label{1.24}
\end{equation}
For relativistic particles as already seen 
\[
\rho \left( T\right) =\frac{\pi ^{2}}{30}g_{*}\left( T\right) \left(
k_{B}T\right) ^{4} 
\]
and 
\[
g_{*}\left( T\right) =\sum_{Bosons}g_{B}+\frac{7}{8}\sum_{F}g_{F} 
\]
denote the effective degrees of freedom. Thus 
\begin{eqnarray}
H &=&\sqrt{\frac{4\pi ^{3}}{45}}\left[ g^{*}\left( T\right) \right] ^{1/2}%
\frac{\left( k_{B}T\right) ^{2}}{\hbar M_{pl}}  \nonumber \\
&=&1.66\left[ g^{*}\left( T\right) \right] ^{1/2}\frac{\left( k_{B}T\right)
^{2}}{\hbar M_{pl}}  \nonumber \\
&=&0.21\left[ g^{*}\left( T\right) \right] ^{1/2}\left( \frac{k_{B}T}{\text{%
MeV}}\right) ^{2}\text{s}^{-1}  \label{1.27}
\end{eqnarray}
where $\hbar M_{pl}=G_{N}^{-1/2}$ is the Planck mass: $M_{pl}\simeq 10^{19}$
GeV.

\subsection{Freeze Out}

At high temperatures $(k_BT\gg m)$, thermodynamic%
\index{cosmology@Cosmology!freeze out@freeze out} equilibrium is maintained
through the processes of decays, inverse decays and scatterings. As the
universe cools and expands, the reaction rates will fail to keep up with the
expansion rate and there will come a time when equilibrium will no longer be
maintained. At various stages then, depending on masses and interaction
strengths, different particles will decouple with a ``freeze out'' surviving
abundance. We now determine conditions under which the statistical
equilibrium is established.

>From dimensional analysis, the reaction rate for a typical process can be
written as follows. For the decay of a $X-$particle, the decay rate is given
by 
\begin{equation}
\gamma _X=g_d\ \alpha _X\ m_X\ 
\frac{m_X}{\left[ \left( k_BT\right) ^2+m_X^2\right] ^{1/2}},  \label{25}
\end{equation}
where $m_X$ is the mass of the $X-$particle, $\alpha _X=\frac{f_X^2}{4\pi }$
is the measure of coupling strength of $X-$particle to the decay products,
and $g_d$ are number of spin states for the decay channels.Note that

\begin{equation}
\begin{tabular}{cccc}
$\gamma _X$ & $\approx $ & $\left\{ 
\begin{array}{c}
g_d\ \alpha _X\ m_X \\ 
g_d\ \alpha _X\ \frac{m_X^2}{k_BT}
\end{array}
\right. $ & $
\begin{array}{c}
k_BT\ll m_X \\ 
k_BT\gg m_X.
\end{array}
$%
\end{tabular}
\end{equation}
The reaction rate for the collision processes is given by 
\[
\gamma _c=\left\langle \sigma \ v\right\rangle \ 
\begin{array}{c}
\lbrack \text{number of target particles per unit volume} \\ 
\text{which is proportional to }\left( k_BT\right) ^3].
\end{array}
\]

Thus 
\begin{eqnarray}
\gamma _{c} &\simeq &\frac{g_{d}\alpha _{X}^{2}\left( k_{B}T\right) ^{5}}{%
\left[ \left( k_{B}T\right) ^{2}+m_{X}^{2}\right] ^{2}}  \label{27} \\
\gamma &\geq &H\,\,\,\,\,\,\,\,\text{Equilibrium} \\
\gamma &\sim &H\,\,\,\,\,\,\,\,\text{Freeze Out} \\
\gamma &\ll &H\,\,\,\,\,\,\,\,\text{Out of Equilibrium}
\end{eqnarray}

\section{Primordial Nucleosynthesis}

After the discovery of cosmic back ground radiation (CMB), the firse success
of big bang cosmology was the correct prediction of primordial abundance of
He ($\simeq 24\%$). This was cooked by nuclear reactions when the universe
was seconds old ($T\simeq 10^{10}$ K$=1$ MeV)

At temperatures $\geq 1$ MeV, the weak reactions such as 
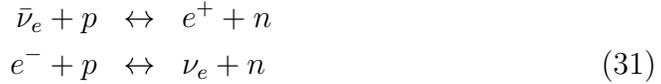
\begin{eqnarray}
\bar{\nu}_{e}+p &\leftrightarrow &e^{+}+n  \nonumber  \label{31} \\
e^{-}+p &\leftrightarrow &\nu _{e}+n  \label{31}
\end{eqnarray}
are still fast compared with the expansion rate of the universe to maintain
thermodynamic equilibrium between $p$ and $n$. The abundance ratio at
equilibrium is given by 
\begin{equation}
\frac{n}{p}\sim e^{-\Delta m/\left( k_{B}T\right) },\qquad
k_{B}T>k_{B}T_{D}\sim 1\ \text{MeV, }t=1\text{ sec}.
\end{equation}
Using $\Delta m=(m_{n}-m_{p})=1.3$ MeV and $k_{B}T=k_{B}T_{D}=1$ MeV, we
find $n/p=0.27$. The rates for the above reactions are given by weak
interactions except that we have to take into account Pauli's exclusion
principle. Then 
\begin{equation}
\gamma ^{\left( n\rightarrow p\right) }=\frac{1}{\pi ^{2}\hbar ^{3}}\frac{%
G_{F}^{2}}{\pi }A\int E_{e}^{2}p_{\nu }^{2}dp_{\nu }\left[ 1+e^{E_{\nu
}/k_{B}T}\right] ^{-1}\left[ 1-\left[ 1+e^{E_{e}/kT}\right] ^{-1}\right]
\end{equation}
where 
\begin{equation}
A=g_{V}^{2}+3g_{A}^{2}=g_{V}^{2}\left( 1+3g_{A}^{2}/g_{V}^{2}\right)
,\,\,\,\,g_{A}/g_{V}\simeq 1.26,g_{V}\simeq 0.9750,
\end{equation}
$g_{V}$ and $g_{A}$ are vector and axial vector coupling constants of the
nucleon. The second factor in the integral is due to Pauli Principle which
suppresses the rate by a factor equal to fraction of all states that are
unfilled. For $k_{B}T\gg Q=\left( m_{n}-m_{p}\right) $ we have 
\[
E_{e}\simeq E_{\nu }\simeq p_{\nu }c=qc=\left( k_{B}T\right) z. 
\]
Thus we obtain 
\[
\gamma ^{\left( n\rightarrow p\right) }\simeq \gamma ^{\left( p\rightarrow
n\right) }\simeq \frac{1}{\pi ^{2}}\frac{G_{F}^{2}}{\pi }A\left( \frac{k_{B}T%
}{\hbar c}\right) ^{3}\left( k_{B}T\right) ^{2}\int_{0}^{\infty
}z^{4}dz\left[ 1+e^{z}\right] ^{-1}\left[ 1+e^{-z}\right] ^{-1} 
\]
The integral can be evaluated by differentiating by parts 
\begin{eqnarray*}
\frac{7}{8}\Gamma \left( 4\right) \zeta \left( 4\right) &=&\int_{0}^{\infty
}z^{3}\frac{dz}{1+e^{z}}=\frac{z^{4}}{4}\left. \left( 1+e^{z}\right)
^{-1}\right| _{0}^{\infty }-\int_{0}^{\infty }\frac{z^{4}}{4}\left(
-1\right) e^{z}\left( 1+e^{z}\right) ^{-2} \\
&=&\frac{1}{4}\int_{0}^{\infty }z^{4}\left[ 1+e^{z}\right] ^{-1}\left[
1+e^{-z}\right] ^{-1}
\end{eqnarray*}
Thus we get 
\begin{equation}
\gamma =\frac{1}{\pi ^{2}}\left( k_{B}T\right) ^{5}\frac{G_{F}^{2}}{\pi }%
A\cdot \frac{7}{2}6\zeta \left( 4\right) =\frac{7\pi }{30}%
G_{F}^{2}g_{V}^{2}\left( 1+g_{A}^{2}\right) \left( k_{B}T\right)
^{5}=0.8\left( \frac{k_{B}T}{\text{MeV}}\right) ^{5}\text{s}^{-1}
\end{equation}

The decoupling temperature is given by 
\begin{eqnarray}
\gamma &=&H \\
&=&0.21g^{*1/2}\left( \frac{k_BT}{\text{MeV}}\right) ^2\text{s}^{-1} 
\nonumber
\end{eqnarray}
where 
\begin{eqnarray*}
g^{*} &=&2\left[ 1+\frac 78\cdot 2+\frac 78\cdot 1\cdot N_\nu \right] \\
&&\,\,\,\,\,\,\,\,\gamma \,\,\,\,\,\,\,\,\,\,\,\,\,e^{\pm
}\,\,\,\,\,\,\,\,\,\,\,\,\,\,\,\,\,\,\,\,\,\nu /\bar{\nu} \\
&=&\frac{22+7N_\nu }4
\end{eqnarray*}
This gives the decoupling temperature 
\begin{eqnarray}
0.8\left( \frac{k_BT_D}{\text{MeV}}\right) ^5 &=&\left( 0.21\right) \left( 
\frac{22+7N_\nu }\Delta \right) ^{1/2}\left( \frac{k_BT_D}{\text{MeV}}%
\right) ^2  \nonumber  \label{37} \\
\frac{k_BT_D}{\text{MeV}} &=&\left[ \frac{0.21}{0.8}\left( \frac{22+7N_\nu }{%
.2}\right) ^{1/2}\right] ^{1/3}  \nonumber \\
&\simeq &1  \label{37}
\end{eqnarray}
if $N_\nu =3.$

As the temperature cools past the decoupling temperature $k_{B}T_{D}\approx
1 $ MeV, it is no longer possible to maintain the thermal equilibrium. The
ratio $n/p$ thereafter is frozen out and is approximately constant (it
decreases slowly due to weak decay of neutron). The freeze out $n/p$ ratio
is given by 
\begin{eqnarray}
X_{n}^{*} &=&n/p\approx e^{-Q/k_{B}T_{D}}\approx 0.16,  \label{38} \\
X_{n}(t) &=&X_{n}^{*}e^{-t/\tau _{n}}  \nonumber
\end{eqnarray}
where $\tau _{n}$ is neutron life time and we have%
\index{cosmology@Cosmology!$Q-$value@$Q-$value} used the $Q-$value $%
Q=(m_{n}-m_{p})+m_{e}=1.8\ $MeV for the reactions (\ref{31}). Helium
nucleosynthesis occurs at $T<T_{S}$ because of deuteron bottle neck. For $%
T>T_{S}$, the deuteron formed is knocked out by photo dissociation 
\[
\gamma +D\rightarrow p+n, 
\]
since the binding energy $\Delta B$ for the deuteron is only 2.2 MeV. The
formation of deuteron actually starts after $k_{B}T_{S}\approx 0.1$ MeV; $%
T_{S}$ is called nucleosynthesis temperature.%
\index{cosmology@Cosmology!nucleosynthesis temperature@nucleosynthesis
temperature} The estimate that $k_{B}T_{S}\approx 0.1$ MeV can be obtained
as follows [$\eta =%
\frac{n_{B0}}{n_{\gamma 0}}$]: 
\begin{equation}
\frac{n_{\gamma }^{\text{diss}}}{n_{B}}\sim \frac{1}{\eta }e^{-\Delta
B/k_{B}T}\leq 1.
\end{equation}
Thus 
\begin{equation}
-\frac{\Delta B}{k_{B}T_{S}}\approx \ln \eta .  \label{40}
\end{equation}
Using $\Delta B\approx 2.2$ MeV$\ $, and $\eta \approx 10^{-10}$, we find $%
k_{B}T_{S}\approx 0.1$ MeV.

For $T>T_S$, photodissociation is so rapid that deuteron abundance is
negligibly small and this provides a bottleneck to further nucleosynethesis.
The deuteron ``bottleneck'' thus delay nucleosynthesis till $k_BT\leq 0.1$
MeV. But once the bottleneck%
\index{cosmology@Cosmology!bottleneck@bottleneck} is passed, nucleosynthesis
proceeds rapidly and essentially all neutrons are incorporated into $%
^{\!4}He:$%
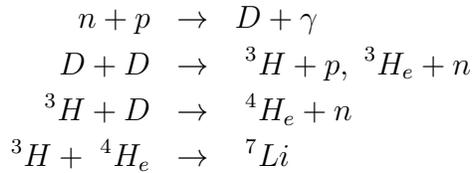
\begin{eqnarray*}
n+p &\rightarrow &D+\gamma \\
D+D &\rightarrow &\ ^3H+p,\ ^3H_e+n \\
^3H+D &\rightarrow &\ ^4H_e+n \\
^3H+\ ^4H_e &\rightarrow &\ ^7Li
\end{eqnarray*}
It is clear from the above reactions that $^4He$ abundance is given by 
\begin{mathletters}
\begin{equation}
Y=%
\frac{2\left( n\ /\ p\right) }{1+n\ /\ p}=\frac{0.32}{1.16}=0.276.
\end{equation}
The ratio $Y$ changes from $T_D$ to $T_S$ due to the neutron decay $%
n\rightarrow p+e^{-}+\bar{\nu}_e$. During this time $n/p$ changes from 0.16
to 0.14. Thus at $T=T_S$, 
\begin{equation}
Y=\frac{0.28}{1.14}=0.246.
\end{equation}

It is clear from Eqs. (\ref{37}) and (\ref{40}) that 
\end{mathletters}
\begin{eqnarray*}
N_\nu &\uparrow &\Longrightarrow T_D\uparrow \Longrightarrow X_n^{*}\uparrow
\Longrightarrow X_n\left( \tau _N\right) \uparrow \Longrightarrow Y\uparrow
\\
\eta &\downarrow &\Longrightarrow T_S\downarrow \Longrightarrow
X_n\downarrow \Longrightarrow Y\downarrow
\end{eqnarray*}
We take $N_\nu =3$ as given by LEP data and use $Y$ to constraint $\eta $.

It turns out that a small amount of $D$ remained unburned. The amount of
unburned $D$ is very sensitive to $\eta $: 
\[
\eta \uparrow \,\,\,\,\,\,\,\,\,\,\,\,\,\,\,\,T_S\uparrow
,\,\,\,\,\,\,\,\,\,\,\,\,\,Y\uparrow \Rightarrow \text{less unburned }D 
\]

Now $D/H$ ratio in primeval samples of the universe has been measured. The
UV\ light (neutral $H$ and $D$ are seen by their UV absorption) came from
distant quasars and absorbers were pregalactic gas clouds. The abundance was
found to be: 
\[
D/H=\left( 3.0\pm 0.1\right) \times 10^{-5}, 
\]
pinning 
\begin{equation}
\eta =\left( 6\pm 3\right) \times 10^{-10}
\end{equation}
We now want to express it in terms of baryon density as a fraction of the
critical density $\rho _{c}$%
\begin{eqnarray*}
\rho _{B} &=&m_{N}n_{B},\,\,\eta =\frac{n_{B_{0}}}{n_{\gamma _{0}}}%
,\,\,n_{\gamma _{0}}=410.50\text{cm}^{-3}\,\, \\
\Omega _{B} &=&\frac{\rho _{B}}{\rho _{c}} \\
\rho _{c} &=&\frac{3H_{0}^{2}}{8\pi G_{N}}=1.054\times 10^{-5}h_{0}^{2}\text{%
GeVcm}^{-3}
\end{eqnarray*}
$H_{0}$ (the present value of Hubble parmater) 
\begin{eqnarray*}
&=&100h_{0}\text{kms}^{-1}\text{Mpc}^{-1} \\
&=&h_{0}\left( 1\times 10^{9}\text{yr}\right) ^{-1} \\
h_{0} &=&0.65\pm 0.05
\end{eqnarray*}
Combining these relations 
\begin{eqnarray}
\Omega _{B}h_{0}^{2} &=&366\times 10^{5}\eta =\left( 0.019\pm 0.001\right) 
\nonumber \\
\Omega _{B} &=&0.045\pm 0.01\text{.}  \label{43}
\end{eqnarray}
This is confirmed by an independent determination of $\Omega _{B}$ involving
measurements of microwave background (CMB) anisotropy, where the underlying
physics is very different, gravitational rather than nuclear. This gives 
\begin{equation}
\Omega _{B}=0.042\pm 0.008
\end{equation}
in very good agreement with the value inferred from Big-Bang Nucleosynthesis
[BBN].

\section{Dark Matter}

At a large scale, measurements of velocity flows of galaxies give 
\begin{equation}
\Omega _m=\frac{\rho _m}{\rho _c}=0.35\pm 0.007
\end{equation}
Such a matter density is much larger than the visible matter density. This
implies that most of the mass in the universe is dark; it does not emit or
absorb any of the electromagnetic ratio. The value of $\Omega _B$ given in
Eq. (\ref{43})is far below the amount of dark matter needed to hold
structures in the universe together. The situation is summarized in the
pyramid shown in Fig. 1.

The detection of the CMB signature of acoustic oscillations in measuerements
of CMB anisotropy also implies 
\begin{eqnarray*}
\Omega _T &=&1.03\pm 0.06 \\
\Omega _k &=&0
\end{eqnarray*}
In the inflationary scienario of the universe, $\Omega $ is driven to unity,
in agreement with the above observed value of $\Omega $. Now from the
relation (\ref{22}) 
\[
H^2=\frac{\dot{R}^2}{R^2}=\frac{8\pi }3G_N\rho -\frac{kc^2}{R^2}+\frac
13\Lambda c^2 
\]
we obtain 
\begin{equation}
1=\Omega _m+\Omega _k+\Omega _\Lambda
\end{equation}
where we have expressed the cosmological constant in terms of vacuum energy $%
\rho _\Lambda :$%
\begin{eqnarray}
\Lambda c^2 &=&4\pi G_N\rho _\Lambda \\
\Omega _\Lambda &=&\frac{\rho _\Lambda }{\rho _c}  \nonumber
\end{eqnarray}
Further $\Omega _k=0$ and $\Omega _m=0.35$ implies that 
\begin{eqnarray}
\Omega _\Lambda &=&1-\Omega _m  \nonumber \\
&=&0.65
\end{eqnarray}
In anology with Dark matter, it is called the Dark energy. Finally 
\begin{equation}
\rho _\Lambda \simeq \frac 23\rho _c\simeq 3\times 10^3\text{eVcm}^{-3}
\label{49}
\end{equation}
giving 
\begin{equation}
\Lambda =4\times 10^{-58}\text{cm}^{-3}  \label{50}
\end{equation}
The nature of both the Dark Matter and the Dark energy is a mystry.

\subsection{Matter-antimatter asymmetry}

We also note that $\eta $ has another role. It appears that the Universe is
matter-antimatter asymmetric. For example anti-proton $\bar{p}$ to proton $p$
ratio in cosmic rays is $\left[ \frac{\bar{p}}{p}\sim 10^{-4}\right] $. In
general $p$ and $\bar{p}$ may annihilate if they are brought togather. In a
typical reaction 
\[
p+\bar{p}\rightarrow \gamma +\gamma . 
\]
The question we wish to answer is, how does the interchange affect the
nuclear density $n_{B}$ or $n_{\bar{B}}.$ We start with $n_{B}=n_{\bar{B}}$.
At $T\leq 1$GeV, the equilibrium abundance of nucleons and antinucleons is,
using Eqs. (\ref{20}) and (\ref{05}) 
\begin{eqnarray}
\eta &=&\frac{n_{B}}{n_{\gamma }}=\frac{n_{\bar{B}}}{n_{\gamma }}  \nonumber
\\
&=&\frac{g_{i}}{2.4}\frac{1}{4}\left( 2\pi \right) ^{1/2}\left( \frac{m_{N}}{%
k_{B}T}\right) ^{3/2}e^{-m_{N}/k_{B}T}  \label{51}
\end{eqnarray}
The freeze out temperature $T$ is given by 
\begin{equation}
\gamma _{ann}\simeq H=1.66\left( g^{*}\right) ^{1/2}\left( k_{B}T^{*}\right)
^{2}\frac{1}{M_{pl}}  \label{52}
\end{equation}
where 
\[
\gamma _{ann}=n_{B}\left\langle \sigma v\right\rangle 
\]
$\sigma $ is the nucleon-antinucleon annhilation crosssection which we may
take as $\frac{1}{m_{\pi }^{2}}$ with $v\simeq c=1$. Thus using Eq. (\ref{20}%
) for $n_{B}$%
\begin{equation}
g_{i}\left( \frac{m_{N}k_{B}T^{*}}{2\pi }\right) ^{3/2}e^{-m_{N}/k_{B}T^{*}}%
\frac{1}{m_{\pi }^{2}}=1.66\left( g^{*}\right) ^{1/2}\left(
k_{B}T^{*}\right) ^{2}\frac{1}{M_{pl}}  \label{53}
\end{equation}
For nucleons $\left( p,n\right) $ and antinucleons, $g_{i}=8$, and 
\[
g^{*}=g_{\gamma }+\frac{7}{8}g_{F}=2+\frac{7}{8}8=9 
\]
Putting $x^{*}=\frac{m_{N}}{k_{B}T^{*}}$ we have 
\begin{eqnarray}
8\frac{1}{\left( 2\pi \right) ^{3/2}}m_{N}x^{*1/2}\frac{1}{m_{\pi }^{2}}%
e^{-x^{*}} &=&1.66\left( 9\right) ^{1/2}\frac{1}{M_{pl}}  \label{54} \\
x^{*-1/2}e^{x^{*}} &=&\frac{8}{3\left( 1.66\right) }\frac{1}{\left( 2\pi
\right) ^{3/2}}\frac{M_{pl}}{m_{\pi }^{2}}m_{N}\simeq 5\times 10^{19}
\label{55}
\end{eqnarray}
Hence $x^{*}=47$. Thus $T^{*}=20$ MeV.

With $T=$ $T^{*}=20$ MeV, $g_{i}=8$, Eq (\ref{51}) gives 
\begin{equation}
\eta =\frac{n_{B}}{n_{\gamma }}=\frac{n_{\bar{B}}}{n_{\gamma }}=2\times
10^{-18}  \label{56}
\end{equation}
This contradicts $\eta =\left( 6\pm 3\right) \times 10^{-10},$ which then
reflect some primordial baryon asymmetry in the universe. To summarize 
\begin{eqnarray}
n_{B} &\rightarrow &n_{B}-n_{\bar{B}}  \nonumber \\
\eta &=&\frac{n_{B}-n_{\bar{B}}}{n_{\gamma }}\simeq \left( 6\pm 3\right)
\times 10^{-10}  \label{57}
\end{eqnarray}

\section{Baryogensis}

\subsection{Sakharov's Conditions}

Towards finding a solution of the second big problem of Cosmology, namly,
that of baryogensis $\left( \eta \simeq 3\times 10^{-10}\right) $,
Sakharov's three conditions, which we enumerate below, must be satisfied:
Assuming that the universe started with a complete matter-antimatter
symmetry in a standard big bang picture, one can obtain matter-antimatter
asymmetry in the universe provided that the following three conditions are
satisfied

(i) Underlying theory must posses 
\[
\Delta B\neq 0 
\]
where $B$ is the baryon number.

(ii) Charge Conjugation, and $CP$ symmetry must be violated; otherwise 
\[
n_{B}\rightarrow _{C,CP}n_{\bar{B}} 
\]
So even if $B$ is violated, one can never establish baryon-antibaryon
asymmetry unless $C$ and $CP$ are violated.

(iii) Departure from thermal equilibrium of X-particles mediating $\Delta
B\neq 0$ processes is necessary. This is because if all processes, including
those which violate baryon number, are in thermal equilibrium, the baryon
asymmetry vanishes. This is a direct consquence of the $CPT$ invariance.

{\bf Proof:}

The density matrix at time $t$ is 
\begin{equation}
\rho \left( t\right) =e^{-\beta \left( t\right) H\left( t\right) },
\label{58}
\end{equation}
where $\beta =\frac 1{k_BT}$. Equilibrium average of $B$ is 
\begin{eqnarray}
\left\langle B\right\rangle _T &=&Tr\left( e^{-\beta H}\hat{B}\right) 
\nonumber \\
&=&Tr\left( \theta ^{-1}\theta e^{-\beta H}\hat{B}\right)  \nonumber \\
&=&Tr\left( \theta e^{-\beta H}\hat{B}\theta ^{-1}\right)  \nonumber \\
&=&Tr\left( \theta e^{-\beta H}\theta ^{-1}\theta \hat{B}\theta ^{-1}\right)
\nonumber \\
&=&Tr\left( e^{-\beta H}\left( -\hat{B}\right) \right)  \nonumber \\
&=&-Tr\left( e^{-\beta H}\hat{B}\right)  \nonumber \\
&=&-\left\langle B\right\rangle _T  \label{59}
\end{eqnarray}
where we have used the fact that $H$ commutes with $CPT\equiv \theta $. Thus 
$\left\langle B\right\rangle _T=0$.

Finally to establish asymmetry dynamically, $B$ violating processes must be
out of equilibrium in the Universe. This can be seen as follows: 
\begin{equation}
\frac{d\Delta n_{B}}{dt}=-\left[ \gamma _{\not{B}}e^{-\left( \frac{m-\mu }{%
k_{B}T}\right) }-\gamma _{\not{B}}e^{-\left( \frac{\bar{m}-\bar{\mu}}{k_{B}T}%
\right) }\right]  \label{60}
\end{equation}
where $\gamma _{\not{B}}$ denotes the rate for $\not{B}$ and $\mu $ is the
chemical potential: $\bar{\mu}=-\mu $. Since $m=\bar{m}$ by $CPT$ theorem, $%
e^{-\frac{m}{k_{B}T}}$ is not relevent and we omit it. Then for $k_{B}T\gg
\mu $, 
\begin{equation}
\frac{d\Delta n_{B}}{dt}=\frac{-2\mu }{k_{B}T}\gamma _{\not{B}}  \label{61}
\end{equation}
On the other hand 
\begin{eqnarray}
\Delta n_{B} &=&\frac{2\zeta \left( 3\right) }{\pi ^{2}}g^{\prime }\left(
k_{B}T\right) ^{3}\left[ e^{\frac{\mu }{k_{B}T}}-e^{\frac{-\mu }{k_{B}T}%
}\right]  \nonumber  \label{62} \\
&\simeq &\frac{2}{\pi ^{2}}g^{\prime }\left( k_{B}T\right) ^{3}\frac{2\mu }{%
k_{B}T}.  \label{62}
\end{eqnarray}
Thus eliminating $\frac{2\mu }{k_{B}T}$%
\begin{eqnarray}
\frac{d\Delta n_{B}}{dt} &=&-\frac{\pi ^{2}}{2}\frac{\gamma _{\not{B}}}{%
g^{\prime }\left( k_{B}T\right) ^{3}}\Delta n_{B}  \nonumber \\
&=&-\frac{\pi ^{2}}{2}\Gamma _{\not{B}}\Delta n_{B}  \label{63}
\end{eqnarray}
where $\Gamma _{\not{B}}=\frac{\gamma _{\not{B}}}{g^{\prime }\left(
k_{B}T\right) ^{3}}=\frac{\gamma _{\not{B}}}{n_{B}}$ gives the rate for $\not%
{B}$. The solution of above equation gives 
\begin{equation}
\Delta n_{B}=\left( \Delta n_{B}\right) _{\text{initial}}e^{-\frac{\pi ^{2}}{%
2}\Gamma _{\not{B}}t}.  \label{64}
\end{equation}
What we learn from this result is that if $B$-violating processes are ever
in equilibrium, then these processes actully washes out any initial
condition for $\Gamma _{\not{B}}t\geq 1.$

After establishing the above preliminaries we shall consider baryogensis at
three levels: (i) Grand Unification (GUT) (ii) Electroweak (iii) Baryogensis
induced by Leptogensis.

\subsection{Baryogensis at GUT Level}

In a typical GUT, quarks and leptons are assigned in one representation so
that the Sakharov condition (i) is naturally satisfied. Taking the example
of the simplest GUT, $SU\left( 5\right) $, the fermions are assigned to the
irreducible representations $\bar{5}_{f}$ and $10_{f}$%
\[
\bar{5}_{f}=\left[ d_{L}^{c},l_{L}\right] ,\,\,\,\,\,10_{f}=\left\{
d_{L},u_{L}^{c},e_{L}^{c}\right\} 
\]
Coupling with gauge bosons is 
\[
\frac{g}{\sqrt{2}}24_{V}\left[ \left( \bar{5}_{f}\right) ^{\dagger }\left( 
\bar{5}_{f}\right) +\left( 10_{f}\right) ^{\dagger }\left( 10_{f}\right)
\right] 
\]
There are 24 gauge bosons [$W^{\pm },Z,\gamma ,8$ gluons and $12$
lepto-quarks $X$, $\bar{X}$ belonging to the adjoint representation $24_{V}$%
].

The condition (iii) of Sakharov is supplied by the expansion of the
Universe. As already mentioned the condition (i) is naturally satisfied in a
GUT, e.g. by lepto-quarks $X$, $\bar{X}$ predicted by GUTs.

At $T=T_{D}$ (the decoupling temperature i.e. the temperature at which $X-$%
particles go out of equilibrium), the number density of $X-$particles is
given by [cf. Eq. (\ref{13})]: 
\begin{equation}
n_{XD}=\frac{2.4}{\pi ^{2}}\frac{g_{X}}{2}\ \left( k_{B}\ T_{D}\right) ^{3},
\end{equation}
where $g_{X}$ is the total number of $X$ (and $\bar{X}$) spin states. Now the%
\index{cosmology@Cosmology!entropy density@entropy density} entropy density
at $T_{D}$ is given by [cf. Eq. (17)] 
\begin{equation}
s=%
\frac{S}{R^{3}}=\frac{4}{3}\rho \left( T\right) =k_{B}\ \frac{\pi ^{2}}{15}%
\left( \frac{4}{3}\right) \ \left( k_{B}\ T_{D}\right) ^{3}g_{*}/2,
\end{equation}
where $g_{*}$ is the effective number of degrees of freedom. The number of
baryons at $T_{D}$ are given by 
\begin{equation}
n_{B}=n_{XD}\ \Delta B.
\end{equation}
where $\Delta B$ denotes baryon asymmetry in $X$-decays. Thus from Eqs. (65)
and (66) 
\begin{eqnarray}
k_{B}\ \left( \frac{n_{B}}{s}\right) _{D} &=&\left( 2.4\right) \frac{45}{%
4\pi ^{4}}\left( \frac{g_{X}}{g_{*}}\right) \ \Delta B  \nonumber \\
&=&0.28\ \left( \frac{g_{X}}{g_{*}}\right) \ \Delta B.
\end{eqnarray}
Now $g_{*}$ is over 100 in a typical GUT. [ In SU(5): $\gamma ,W^{\pm
},Z^{0},$8G's, 34 Higgs, 6 quarks, 3 leptons, 3 neutrinos, 12 $X$'s. Thus $%
g_{*}=(24\times 2)+34+\frac{7}{8}(18\times 4+3\times 4+3\times 2)=160.8$].
We, therefore, expect $g_{X}/g_{*}\approx 10^{-2}$ to $10^{-1}$. Thus we
have 
\begin{equation}
k_{B}\ \left( \frac{n_{B}}{s}\right) _{D}\approx 0.28\times \left( 10^{-2}%
\text{ }-10^{-1}\right) \ \Delta B_{X}\approx 3\times \left(
10^{-3}-10^{-2}\right) \ \Delta B.
\end{equation}
But $\left( n_{B}/s\right) _{D}$ $=\left( n_{B}/s\right) _{0}$, where $0$
denotes the present time. Thus 
\begin{equation}
k_{B}\ \left( \frac{n_{B}}{s}\right) _{0}\approx 3\times \left( 10^{-3}\text{
}-10^{-2}\right) \ \Delta B.
\end{equation}
\begin{eqnarray}
\left( \frac{s}{k_{B}}\right) _{0} &=&\left( \frac{s}{k_{B}}\right) _{\gamma
_{0}}+\left( \frac{s}{k_{B}}\right) _{\nu _{0}}  \nonumber \\
&=&\frac{2\pi ^{4}}{45\times 1.2\text{ }}\left[ \frac{1}{2}\ g_{\gamma }\
n_{\gamma _{0}}+\frac{1}{2}\times \frac{4}{3}\ \sum_{i}g_{\nu i}\ n_{\nu
_{0}}\right]  \nonumber \\
&=&3.6\left[ 1+\frac{21}{4}\frac{1}{2}\frac{4}{11}\right] \ n_{\gamma
_{0}}\approx 7\ n_{\gamma _{0}}.  \label{71}
\end{eqnarray}
Now [cf. Eqs. (17) and (5)] where we have used that $n_{\nu _{0}}=\left(
3/11\right) n_{\gamma _{0}}$ and $\sum_{i}g_{\nu i}=3(7/8).1.2.$ Hence from
Eq. (70), we get 
\begin{eqnarray}
\left( \frac{n_{B}}{n_{\gamma }}\right) _{0} &\approx &21\times \left(
10^{-3}\text{ to }10^{-2}\right) \ \Delta B\approx 2\times \left( 10^{-2}%
\text{ to }10^{-1}\right) \ \Delta B.  \nonumber \\
&\simeq &A\left( \Delta B\right) ,
\end{eqnarray}
where $A\sim 10^{-1}-10^{-2}.$ Now we can write 
\begin{eqnarray}
\Delta B &=&\sum_{f}B_{f}\frac{\left[ \Gamma \left( X\rightarrow f\right) -%
\bar{\Gamma}\left( \bar{X}\rightarrow \bar{f}\right) \right] }{\Gamma
_{tot}\left( X\right) }, \\
f &\equiv &\left\{ ql,\bar{q}\bar{q}\right\} ,\,\,B_{f}=\frac{1}{3},-\frac{2%
}{3}  \nonumber
\end{eqnarray}
$\Delta B$ vanishes if CP and C are conserved. The $X-$particles can
generate $\Delta B$, by the processes of the following type [$r$ is the
branching ratio] 
\begin{eqnarray*}
X &\rightarrow &ql:r\qquad B_{1}=1/3 \\
X &\rightarrow &\bar{q}\bar{q}:1-r\qquad \bar{B}_{2}=-2/3 \\
\bar{X} &\rightarrow &\bar{q}\bar{l}:\bar{r}\qquad \bar{B}_{1}=-1/3 \\
\bar{X} &\rightarrow &qq:1-\bar{r}\qquad B_{2}=2/3.
\end{eqnarray*}
The mean baryon number per decay 
\begin{eqnarray}
B_{X} &=&r\ B_{1}+\left( 1-r\right) \bar{B}_{2}  \nonumber \\
B_{\bar{X}} &=&\bar{r}\ \bar{B}_{1}+\left( 1-\bar{r}\right) B_{2}.
\end{eqnarray}
Thus 
\begin{eqnarray}
\Delta B &=&\frac{1}{2}\left[ r\ B_{1}+\left( 1-r\right) \bar{B}_{2}+\bar{r}%
\ \bar{B}_{1}+\left( 1-\bar{r}\right) B_{2}\right]  \nonumber \\
&=&\frac{1}{2}\left[ r\ \left( B_{1}-\bar{B}_{2}\right) +\bar{r}\ \left( 
\bar{B}_{1}-B_{2}\right) +\left( \bar{B}_{2}+B_{2}\right) \right]  \nonumber
\\
&=&\frac{1}{2}\left( r\ -\bar{r}\right) .
\end{eqnarray}
>From Eqs. (75) and (72), we see that we can explain the baryon number
generation if $r\neq \bar{r}$ i.e. $X-$interactions violate $C$ and CP. Also
we require $\Delta B\sim 10^{-8}$ in order to explain the present baryon
number $\eta =n_{B}/n_{\gamma }\approx 10^{-10}$.

Let us now obtain an estimate for $T_{D}$. If $k_{B}\ T_{D}>m_{X}$, the
thermal equilibrium can be maintained by inverse decays. Thus the condition
for departure from equilibrium is [cf. Eqs. (24) and (26) $k_{B}T_{D}\simeq
m_{X}$]: 
\begin{equation}
\frac{1}{3}\alpha _{X}\ g_{d}\ \left( k_{B}\ T_{D}\right) \approx 1.66\
g_{*}^{1/2}\ \frac{\left( k_{B}\ T_{D}\right) ^{2}}{M_{pl}}.
\end{equation}
The factor 1/3 is due to spin average [$X$ is a vector particle]. Now using $%
g_{d}\approx 12\times 2=24$ and $g_{*}\approx 160$, we get 
\begin{equation}
k_{B}\ T_{D}\approx \alpha _{X}\ \left( 4.0\right) 10^{18}GeV.
\end{equation}
Using $\alpha _{X}\approx 1/40$ [SU(5) value], we get 
\begin{equation}
k_{B}\ T_{D}\approx 10^{17}GeV.
\end{equation}
Thus if $X-$bosons are vector bosons, $k_{B}\ T_{D}>$ mass of vector bosons
of SU(5) [$\simeq 10^{15}-10^{16}$ GeV]and therefore vector bosons of SU(5)
cannot give rise to baryon asymmetry.

\subsection{Can Higgs particles give rise to the baryon asymmetry?}

Whereas the gauge sector of the $SU(5)$ structure is uniquely determined by
the gauge group, in the Higgs sector the results depend on the choice of
representation. Now the Higgs fields that couple to fermions belong to 
\[
5_H,10_H,15_H,45_H\text{ and }50_H. 
\]
Let us first consider the minimal $5_H$ which contains Higgs doublet of the
SM:

Color singlet 
\[
\left( 1,2,1/2\right) ,\,\,\,\,\,\,\,\,\,\,\, 
\]
the latter two quantum numbers refer to 
\[
SU(2)\times U(1). 
\]
$B$ violating color triplet 
\[
H_3\,\,\left( 3,1,-1/3\right) . 
\]
At tree level, the Higgs coupling with fermions are shown in Fig. 2. In any
vertex of $SU(5)$, $B-L$ is preserved.

$h_U$, $h_D$ are complex Yukawa coupling matrices. CP--violation arises from
complex phases of the Yukawa couplings, which can not be absorbed by field
redefinations. At tree level, these phases do not give any contribution to
baryon asymmetry since $\func{Im}$ $Tr\left[ h_D^{\dagger }h_D\right] =0$.
One has to go to loop level, where denoting by $\chi $, a member of $H_3$,
where $\phi $ is some exchanged state, another Higgs or a gauge boson. Thus $%
r$ is given by Fig. 3. 
\begin{equation}
r\sim \left| \gamma _0+\gamma _1I\left( M_\chi ^2-i\varepsilon \right)
\right| ^2
\end{equation}
where $\gamma _0$ and $\gamma _1$ are complex and $I$ has an analytical
structure. Then 
\begin{eqnarray}
\frac 12\left( r-\bar{r}\right) &\sim &\frac 12\left\{ \left| \gamma
_0+\gamma _1I\left( M_\chi ^2-i\varepsilon \right) \right| ^2-\left| \gamma
_0^{*}+\gamma _1^{*}I\left( M_\chi ^2-i\varepsilon \right) \right| ^2\right\}
\nonumber \\
&\sim &2\func{Im}\left( \gamma _0\gamma _1^{*}\right) \func{Im}\left(
I\left( M_\chi ^2-i\varepsilon \right) \right)
\end{eqnarray}
where the first part is the $CP$ violating and the second part is determined
by the rescattering dynamics. Thus 
\begin{equation}
\eta =A\Delta B_\chi =A\frac 12\left( r-\bar{r}\right) \approx A\func{Im}%
\left( \gamma _0\gamma _1^{*}\right) \func{Im}I
\end{equation}
where $A$ comes from out of the equilibrium condition as seen previously; $%
\func{Im}\left( \gamma _0\gamma _1^{*}\right) $ gives $CP$ and $C$ violation
while $\func{Im}I$ comes from GUT's dynamics. There is no firm pediction for 
$\eta .$

In particular, take $\phi $ in the above figure as $5_H$ as shown in Fig.4.
We can choose $f$ to be real and $h$ has 3 phases. Even so we can not
generate in the lowest non-trival order CP violation. This is because $%
\gamma _0\sim f$ real and $\gamma _1\sim fh\bar{h}$, which has no $CP$
phase. Also gauge exchange [$\phi \equiv G^\mu $] give no phase. One can
eventually generate $\eta $ to higher order loop graphs. But then $\eta \sim
10^{-16}$, which is too small .

It is possible to obtain $\eta \sim 10^{-10}$ by either (i) adding Higgs in
the $45$ representation or (ii) by using more elaborate GUT's e.g. $SO(10)$.
In $SO(10)$ there exists a fermion that is singlet under SM, carries $L=-1$
and is identified with $\nu _{R}$. $CP$ violation may be provided by the
complex Yukawa couplings between the right-handed and the left-handed
neutrinos and scalar Higgs. The right-handed neutrinos acquire a Majorana
mass $M_{N}=O\left( B-L\right) $ i.e. at the scale where $U(1)_{B-L}$ is
broken, and its out of equilibrium decays may generate a non-vanishing $%
\left( B-L\right) $ asymmetry. We shall come back to the role of
right-handed neutrino in generating $\eta $ when we consider Leptogenesis.

\section{Electroweak Baryogenesis}

In the SM, both baryon number ,$B$, and lepton number ,$L$, symmetries hold
at the classical level 
\[
{\cal L}_{\text{SM}}\rightarrow _{B,L}{\cal L}_{\text{SM}}. 
\]
However, because of the chiral nature of electroweak theory, at quantum
level both $J_{B}^{\mu }$ and $J_{L}^{\mu }$ are not conserved, so called
electroweak anamoly: 
\begin{eqnarray}
\partial _{\mu }J_{B}^{\mu } &=&\partial _{\mu }J_{L}^{\mu }  \nonumber \\
&=&N_{g}\left( \frac{\alpha _{2}}{\pi }W_{a}^{\mu \nu }\tilde{W}_{a\mu \nu }-%
\frac{\alpha ^{\prime }}{8\pi }F_{\mu \nu }\tilde{F}_{\mu \nu }\right)
\end{eqnarray}
where $N_{g}$ is the number of generations, $\alpha _{2}=\frac{g_{2}^{2}}{%
4\pi },$ $\alpha ^{\prime }=\frac{g^{\prime 2}}{4\pi }$ are couplings
corresponding to $SU_{L}(2)\times U(1)$; $\tilde{W}_{\mu \nu }=\frac{1}{2}%
\varepsilon ^{\mu \nu \alpha \beta }W_{\alpha \beta }$. Then 
\begin{eqnarray}
\partial _{\mu }\left( J_{B}^{\mu }-J_{L}^{\mu }\right) &=&\partial _{\mu
}J_{B-L}^{\mu }=0  \nonumber \\
\partial _{\mu }\left( J_{B}^{\mu }+J_{L}^{\mu }\right) &\neq &0
\end{eqnarray}
Further 
\begin{eqnarray}
\Delta B &=&B\left( +\infty \right) -B\left( -\infty \right)  \nonumber \\
&=&\int_{-\infty }^{\infty }dt\partial _{0}\int d^{3}xJ_{B}^{0}\left( \vec{x}%
,t\right)  \nonumber \\
&=&\int_{-\infty }^{\infty }dt\int d^{3}x\partial _{\mu }J_{B}^{\mu }\left( 
\vec{x},t\right)
\end{eqnarray}
since by Gauss's theorm we can convert $\int d^{3}x\partial
_{i}J_{B}^{i}\left( \vec{x},t\right) $ into a surface integral which we can
put equal to zero. Similarly for $\Delta L$. Note that $\Delta \left(
B-L\right) =0$; the electroweak anamoly preserves $B-L$. But 
\begin{eqnarray}
\Delta \left( B+L\right) &\neq &0  \nonumber \\
&=&2N_{g}\nu
\end{eqnarray}
where 
\[
\nu =\frac{\alpha _{2}}{8\pi }\int d^{4}xW_{a}^{\mu \nu }\tilde{W}_{a\mu \nu
} 
\]
Now in some theories, one can find classical solutions of the Euclidean
field equations. These solutions, called instantons, are localized in
Euclidean time as well as in space. Let us consider a pure Yang-Mills field:
Time $t$ in Minkowsky space must be replaced by $it$ in Euclidean space.
Then the Euclidean Action 
\begin{eqnarray}
S_{E} &=&\int d^{4}x_{E}{\cal L}_{E}\left( x_{E}\right) =-iS  \nonumber \\
{\cal L}_{E}\left( x_{E}\right) &=&\frac{1}{g^{2}}F_{\mu \nu }^{a}F_{\mu \nu
}^{a}
\end{eqnarray}
There is no distinction between covariant and contravariant indices in
Euclidean space. Here 
\begin{eqnarray*}
A_{\mu }^{a} &\rightarrow &\frac{i}{g}A_{\mu }^{a} \\
F_{\mu \nu }^{a} &\rightarrow &\frac{i}{g}F_{\mu \nu }^{a} \\
F_{\mu \nu }^{a} &=&\partial _{\mu }A_{\nu }^{a}-\partial _{\nu }A_{\mu
}^{a}+\left[ A_{\mu }^{a},A_{\nu }^{a}\right] \\
A_{\mu } &=&\sum_{a}T^{a}A_{\mu }^{a} \\
F_{\mu \nu } &=&\sum_{a}T^{a}F_{\mu \nu }^{a}
\end{eqnarray*}
It can be shown that 
\begin{eqnarray}
S_{E}\left[ A\right] &\geq &\frac{1}{2}\int d^{4}xTr[F_{\mu \nu }\tilde{F}%
_{\mu \nu }]  \nonumber \\
&=&8\frac{\pi ^{2}}{g}\left| \nu \left[ A\right] \right| ,
\end{eqnarray}
the lower bound is obtained when 
\[
\tilde{F}_{\mu \nu }=\pm F_{\mu \nu }. 
\]

It is necessary to find an interpretation for instanton solution in
Euclidean four-space (imaginary $t$) within Minkowsky $3+1$ space-time (time
real) in order to understand their physical significance. It has been shown
that instantons become `tunneling events' between two different Minkowski
events. The gauge vacuum is given by 
\begin{equation}
\left| \theta \right\rangle =\sum_ie^{-in\theta }\left| n\right\rangle .
\end{equation}
It has been shown by t'Hooft that the transition probility between two
closet vacuua is given by 
\begin{equation}
\sim e^{-8\pi ^2/g^2}\,\,\,\,\,\,\,\,\,\,\text{for }\nu =1.
\end{equation}
This is the zero temperature solution: 
\begin{equation}
S_E\sim \frac{8\pi ^2}{g^2}
\end{equation}
For our problem $\frac{g^2}{4\pi }=\alpha _2$. Thus $\left( B+L\right) $
violating amplitude is 
\begin{equation}
A\sim e^{\left( -2\pi /\alpha _2\right) \nu }
\end{equation}
which is extremely small$\sim 10^{-90}$. Thus the probability of
B--violating processes is highly supressed at zero temperature.

The situation is different at high temperature which is relevent at early
Universe. Here one goes from one vacuum to another by thermal flucatuations
rather than tunneling [see Fig. 5].

The transition probability (per unit time per unit volume) is given by 
\begin{equation}
P\sim e^{-V_{0}(T)/T}
\end{equation}
where $V_{0}(T)$ is the height of the barrier. If the system is able to
perform a transition from one vacuum to the closest one, then $\Delta \left(
B+L\right) =2N_{g}=6.$

Each transition creats nine left-handed quarks as only these are coupled to $%
W$ bosons (3 color states for each generation) and left-handed leptons (one
per generation). However, adjacent vacuua of the EW theory are separated by
a ridge of configrations with energies larger than that of the vacuua. The
lowest energy point on this ridge is a saddle point solution to the
equations of motion, and is refered to as the Sphaleron.

The thermal rate of B--violation in the broken phase is proportional to [in
units $k_B=1$] 
\[
\exp (-S_3/T) 
\]
where $S_3$ is the three dimentional action computed along the sphaleron
configration: 
\begin{equation}
S_3\equiv E_{sp}(T)\equiv C\left( \frac{m_H}{m_W}\right) \frac{\pi m_W\left(
T\right) }{\alpha _2}
\end{equation}
where $C\left( \frac{m_H}{m_W}\right) $ is a function of $\lambda $, $\frac{%
m_H^2}{m_W^2}\sim \lambda $: 
\[
E_{sp}\simeq 7-14\text{ TeV as }\lambda \text{ increases from }0\text{ to }%
\infty . 
\]
Then, the transition probability per unit time per unit volume is 
\begin{equation}
{\cal P}_{sp}\left( T\right) =\mu \left[ \frac{m_W}{\alpha _2T}\right]
^3m_W^4\exp \left[ \frac{-E_{sp}(T)}T\right]
\end{equation}
where $\mu $ is a dimensionless constant and the Boltzmann suppression
appears large.

However, it is to be expected that, when EW symmetry becomes restored at
temperature of around $100$ GeV [$m_W(T)\rightarrow 0$, Electroweak phase
transition] there will no longer be an exponential suppression. Now the only
important scale in the symmetric phase is $\alpha _2T$ so that dimensional
ground, we expect 
\begin{equation}
{\cal P}_{sp}\left( T\right) ={\cal K}\left( \alpha _2T\right) ^4
\end{equation}
where numerical estimates yielded ${\cal K\sim }0.1-1.$

Lattice simulation indicates 
\begin{equation}
{\cal P}_{sp}\sim 30\alpha _{2}^{5}T^{4}
\end{equation}
not very different from above as $\alpha _{2}\sim \frac{1}{30}.$

Now $\not{B}$ and $\not{L}$ processes are in thermal equilibrium for 
\begin{eqnarray}
\Gamma _{sp} &=&\frac{{\cal P}_{sp}}{T^3}>H\simeq 1.66g^{*\,1/2}\frac{T^2}{%
M_{\text{pl}}} \\
{\cal K}\alpha _2^4T &>&1.66g^{*\,1/2}\frac{T^2}{M_{\text{pl}}}  \nonumber \\
\alpha _2 &=&\frac{\alpha _e}{\sin ^2\theta _W}=\frac 4{137}\simeq 0.029 
\nonumber \\
\alpha _2^4 &\simeq &10^{-6} \\
T &<&\left( \frac{{\cal K}}{1.66g^{*\,1/2}}\right) \alpha _2^4M_{\text{pl}%
}\approx 10^{12}\text{ GeV}
\end{eqnarray}
Thus $\not{B}$ and $\not{L}$ processes are in thermal equilibrium for the
temperature in the range 
\begin{equation}
T_{\text{EW}}\simeq 100\text{ GeV}<T<T_{sp}\simeq 10^{12}\text{ GeV}
\end{equation}
This implies any $\Delta n_{B+L}\,$ established above 
\[
T_{\text{max}}\left( \sim \alpha _2^4M_{\text{pl}}\right) \text{ } 
\]
e.g., GUT, will get washed out down to $T_{\text{EW}}.$

Given the above result, to generate $\eta $ at the EW phase transition:
First the baryon number is violated. as we have seen above. Second $CP$ is
violated, in the standard model. The third condition of Sakharov can be
satisfied if the EW transition is of first order since then the coexistance
of broken and unbroken phases at the phase transition is a departure from
equilibrium. However one can not get the first order transition unless Higgs
is light, $m_{H}<60$ GeV, which is ruled out by LEP, $m_{H}\geq 114$ GeV.
Another problem is the size of CP--violation: 
\begin{equation}
\eta \simeq \alpha _{2}^{4}\epsilon _{C\not{P}}\simeq 10^{-6}\epsilon _{C\not%
{P}}
\end{equation}

where 
\begin{equation}
\epsilon _{C\not{P}}\sim \lambda ^6\sin \delta
\end{equation}
due to several GIM suppression factors in CKM, $\lambda \simeq 0.22$ so that 
\begin{equation}
\lambda ^6\simeq 5.5\times 10^{-6}.
\end{equation}
This gives 
\[
\eta \sim 5.5\times 10^{-12}\delta \sim 10^{-18} 
\]

The other possibility is the leptogensis, where one tries to generate $L\neq
0$ but no B from neutrino physics well before the electroweak transition,
and L gets partially converted into B due to electroweak anamoly. This is
discussed in the next section.

\section{Baryogenesis via Leptogenesis}

As already seen sphaleron transitions lead to 
\begin{eqnarray*}
\Delta \left( B-L\right) &=&0 \\
\Delta \left( B+L\right) &=&2N_{g}=6
\end{eqnarray*}
the baryon asymmetry can be generated by the lepton asymmetry.

Further $B+L$ asymmetry generated before EW transition i.e. at $T>T_{\text{EW%
}},$ will be washed out. However, since only left handed fields couple to
sphalerons, a non zero value of $B+L$ can persist in the high temperature
symmetric phase if there exist a non vanishing $B-L$ asymmetry [see below].
As already seen 
\[
n_i-\bar{n}_i=\frac 2{\pi ^2}g^{\prime }T^3\left( \frac{2\mu _i}T\right) 
\]
This also implies 
\begin{eqnarray}
n_B &=&B\left( \frac 4{\pi ^2}g^{\prime }T^2\right)  \nonumber \\
n_L &=&L\left( \frac 4{\pi ^2}g^{\prime }T^2\right)  \label{104}
\end{eqnarray}
where $B$ and $L$ are baryon and lepton asymmetry respectively.

Note that in SM 
\begin{eqnarray*}
q_{Li} &=&\left( 
\begin{array}{c}
u_{Li} \\ 
d_{Li}
\end{array}
\right) \text{ \thinspace \thinspace \thinspace \thinspace \thinspace
\thinspace \thinspace }B=\frac{1}{3}\text{, }L=0 \\
&&u_{Ri},d_{Ri} \\
\ell _{Li} &=&\left( 
\begin{array}{c}
\nu _{Li} \\ 
e_{Li}
\end{array}
\right) \text{ \thinspace \thinspace \thinspace \thinspace \thinspace
\thinspace \thinspace }B=0\text{, }L=1 \\
&&\nu _{Ri},e_{Ri}
\end{eqnarray*}
Thus in Eq. (\ref{104}) 
\begin{eqnarray}
B &=&3\times \frac{1}{3}\sum_{i}\left( 2\mu _{qi}+2\mu _{ui}+2\mu
_{di}\right)  \nonumber \\
L &=&\sum_{i}\left( 2\mu _{li}+2\mu _{ei}\right)  \label{105}
\end{eqnarray}
In high temperature plasma quarks, leptons and Higgs interact via Yukawa and
gauge couplings and in addition, via the non perturbative sphaleron
processes. In thermal equilibrium all these processes yield constraints
between various chemical potentials. The effective interaction 
\[
O_{B+L}=\Pi _{i}\left( q_{Li}q_{Li}q_{Li}\ell _{Li}\right) 
\]
yields 
\begin{equation}
\sum_{i}\left( 3\mu _{qi}+\mu _{li}\right) =0  \label{106}
\end{equation}
Another constraint is provided by vanishing of total charge of plasma 
\[
\sum_{i}\left[ 
\begin{array}{c}
3\frac{1}{3}2\mu _{qi}+3\frac{4}{3}\mu _{ui} \\ 
+3\left( -\frac{2}{3}\right) \mu _{di}+\left( -1\right) 2\mu _{li}+\left(
-2\right) \mu _{ei}+\frac{1}{N}\left( 1\right) \mu _{\phi }
\end{array}
\right] =0 
\]
where we have used 
\[
Y_{q}=\frac{1}{3},\,Y_{u}=\frac{4}{3},\,Y_{d}=-\frac{2}{3}%
,\,Y_{l}=-1,\,Y_{e^{-}}=-2,\,Y_{\phi }=1 
\]
The above equation can be written as 
\begin{equation}
\sum_{i}\left( \mu _{qi}+2\mu _{ui}-\mu _{di}-\mu _{li}-\mu _{ei}+\frac{2}{N}%
\mu _{\phi }\right) =0  \label{107}
\end{equation}
Further invariance of Yukawa couplings $\bar{q}_{Li}\phi d_{Ri}$, etc give 
\begin{eqnarray}
\mu _{qi}-\mu _{\phi }-\mu _{dj} &=&0  \nonumber  \label{108} \\
\mu _{qi}-\mu _{\phi }-\mu _{uj} &=&0  \nonumber \\
\mu _{li}-\mu _{\phi }-\mu _{ej} &=&0  \label{108}
\end{eqnarray}
When all Yukawa interactions are in equilibrium, these interactions
establish equilibrium in different generations 
\[
\mu _{li}=\mu _{l},\,\mu _{qi}=\mu _{q}\text{ etc.} 
\]
Thus we obtain from Eqs (\ref{106}) and (\ref{107}) 
\begin{eqnarray*}
\mu _{q} &=&-\frac{1}{3}\mu _{l} \\
\mu _{q}+2\mu _{u}-\mu _{d}-\mu _{l}-\mu _{e}+\frac{2}{N}\mu _{\phi } &=&0
\end{eqnarray*}
giving 
\begin{equation}
-\frac{4}{3}\mu _{l}+2\mu _{u}-\mu _{d}-\mu _{e}+\frac{2}{N}\mu _{\phi }=0
\label{109}
\end{equation}
Further Eqs. (\ref{108}) imply 
\begin{eqnarray}
-\frac{1}{3}\mu _{l}-\mu _{\phi }-\mu _{d} &=&0  \nonumber \\
-\frac{1}{3}\mu _{l}-\mu _{\phi }-\mu _{u} &=&0  \nonumber \\
\mu _{l}-\mu _{\phi }-\mu _{e} &=&0  \label{110}
\end{eqnarray}
Using the above equations, we can write (\ref{109}) as 
\[
-\frac{4}{3}\mu _{l}+2\left( -\frac{1}{3}\mu _{l}+\mu _{\phi }\right)
-\left( -\frac{1}{3}\mu _{l}-\mu _{\phi }\right) -\left( -\mu _{l}-\mu
_{\phi }\right) +\frac{2}{N}\mu _{\phi }=0 
\]
Thus finally we can express $\mu _{q}$, $\mu _{u}$, $\mu _{d}$, $\mu _{e}$,
and $\mu _{\phi }$ interms of $\mu _{l}.$%
\begin{eqnarray}
\mu _{\phi } &=&\frac{8}{3}N\frac{1}{4N+2}\mu _{l}=\frac{4N}{6N+3}\mu _{l} 
\nonumber \\
\mu _{d} &=&-\frac{1}{3}\mu _{l}-\mu _{\phi }  \nonumber \\
&=&-\frac{1}{3}\mu _{l}-\frac{4N}{6N+3}\mu _{l}  \nonumber \\
&=&-\frac{6N+1}{6N+3}\mu _{l}  \nonumber \\
\mu _{u} &=&-\frac{1}{3}\mu _{l}+\mu _{\phi }  \nonumber \\
&=&-\frac{1}{3}\mu _{l}+\frac{4N}{6N+3}\mu _{l}  \nonumber \\
&=&\frac{2N-1}{6N+3}\mu _{l}  \nonumber \\
\mu _{e} &=&\mu _{l}-\mu _{\phi }  \nonumber \\
&=&\mu _{l}-\frac{4N}{6N+3}\mu _{l}  \nonumber \\
&=&\frac{2N+3}{6N+3}\mu _{l}
\end{eqnarray}
Hence from Eqs. (\ref{105}) 
\begin{eqnarray}
B &=&N\left\{ -\frac{2}{3}\mu _{l}+\frac{2N-1}{6N+3}\mu _{l}-\frac{6N+1}{6N+3%
}\mu _{l}\right\}  \nonumber \\
&=&\left[ -4N-2+2N-1-6N-1\right] \frac{\mu _{l}}{6N+3}  \nonumber \\
&=&-N\frac{\left( 8N+4\right) }{3\left( 2N+1\right) }\mu _{l}  \nonumber \\
&=&-\frac{4N}{3}\mu _{l} \\
L &=&N\left( 2\mu _{l}+\frac{2N+3}{6N+3}\mu _{l}\right)  \nonumber \\
&=&\frac{14N^{2}+9N}{6N+3}\mu _{l} \\
B-L &=&-\frac{8N^{2}+4N+14N^{2}+9N}{6N+3}\mu _{l}  \nonumber \\
&=&-\frac{22N^{2}+13N}{6N+3}\mu _{l} \\
\frac{B}{B-L} &=&\frac{8N^{2}+4N}{22N^{2}+13N}  \nonumber \\
&=&\frac{8N+4}{22N+13}  \nonumber \\
&=&\frac{8N_{g}+4n_{H}}{22N_{g}+13n_{H}}\equiv a
\end{eqnarray}
These relations hold for $T\gg v$. In general $B/B-L$ is a function of $v/T$%
. For SM, $N_{g}=3$, $n_{H}=1$ so that $a=28/79$.

Thus finally we obtain 
\begin{eqnarray}
Y_{B}( &\equiv &\frac{n_{B}-n_{\bar{B}}}{s})  \nonumber \\
&=&aY_{B-L}=\frac{a}{a-1}Y_{L}  \label{116}
\end{eqnarray}
Note that by using Eq. (\ref{71}), 
\begin{eqnarray*}
Y_{B} &=&\eta \left( \frac{\eta _{\gamma }}{s}\right) \simeq \frac{1}{7}\eta
\\
&&\frac{1}{7}\left( 6\pm 3\right) \times 10^{-10}
\end{eqnarray*}

In SM as well as in $SU(5)$, $B-L$ is conserved and no asymmetry in $B-L$
can be generated. However, adding a right handed Majorana neutrino to the SM
breaks $B-L$, and the primordial lepton asymmetry may be generated by the
out of equilibrium decay of heavy right handed Majorana neutrino $N_R$. The
simple extension of SM can be embedded in GUTs with gauge group containing $%
SO(10)$. Majorana neutrinos can also lead to See-saw mechanism, explaining
the smallness of light neutrino $\nu $ masses.

The relevent couplings are 
\begin{equation}
{\cal L}=\bar{\ell}_L\phi h_\nu N_R+\frac 12\bar{N}_R^cMN_R+h.c.  \label{117}
\end{equation}
where $\phi $ is the usual Higgs doublet under $SU(2)_L$ while the second
term gives Majorana mass for the right handed neutrino $N$. The vacuum
expection value of the Higgs field $\left\langle \phi \right\rangle $
generates neutrino Dirac masses 
\begin{equation}
m_D=h_\nu \left\langle \phi \right\rangle  \label{118}
\end{equation}
The Lagrangian (\ref{117}) also generate an effective $\Delta L=2$ Lagrangian

\begin{eqnarray}
&&{\cal L}_{\Delta L=2}  \nonumber \\
&=&\frac GM\left( \ell _L^Ti\sigma _2\phi \right) C^{-1}\left( \phi
^Ti\sigma _2\ell _L\right)  \label{119}
\end{eqnarray}
This generates a Majorana mass for light neutrinos 
\begin{equation}
m_\nu =\frac GM\left\langle \phi \right\rangle _0^2  \label{120}
\end{equation}
Further 
\begin{equation}
m_N\equiv M\gg m_D
\end{equation}

If the $\Delta L=2$ interactions are in equilibrium, but the right handed
electrons are not, then $\mu _l-\mu _\phi -\mu _e=0$ is replaced by [c.f.
Eq. (\ref{117})] 
\begin{equation}
\mu _l+\mu _\phi =0\Rightarrow \mu _l=-\mu _\phi  \label{121}
\end{equation}
Thus using all the other previous equations 
\begin{eqnarray*}
\mu _q &=&\frac 13\mu _l \\
\mu _d &=&\frac 23\mu _l \\
\mu _u &=&-\frac 43\mu _l \\
\mu _\phi &=&-\mu _l
\end{eqnarray*}
\begin{equation}
\mu _q+2\mu _u-\mu _d-\mu _l+\frac 2N\mu _\phi =0  \label{122}
\end{equation}
The equations give 
\begin{eqnarray}
\mu _l &=&-\frac{3N}{14N+6}\mu _e  \nonumber \\
\mu _d &=&-\frac{2N}{14N+6}\mu _l  \nonumber \\
\mu _u &=&\frac{4N}{14N+6}  \nonumber \\
\mu _q &=&\frac N{14N+6},  \label{123}
\end{eqnarray}
so that from Eq. (\ref{105}) 
\begin{eqnarray}
B &=&N\left[ \frac{2N}{14N+6}+\frac{4N}{14N+6}-\frac{2N}{14N+6}\right] \mu _e
\nonumber \\
&=&\frac{4N^2}{14N+6}\mu _e  \nonumber \\
L &=&N\left[ -\frac{6N}{14N+6}+1\right] \mu _e  \nonumber \\
&=&\frac{8N^2+6N}{14N+6}\mu _e  \nonumber \\
B-L &=&\frac{-4N^2-6N}{14N+6}  \nonumber \\
\frac B{B-L} &=&\frac{4N^2}{-4N^2-6N}=\frac{-2N}{2N+3}=a;\,\,\,a-1=\frac{%
-4N-3}{2N+3}  \nonumber \\
\frac BL &=&\frac{2N}{4N+3}=\frac a{a-1}  \label{124}
\end{eqnarray}
The above relations hold if the corresponding interactions are in thermal
equilibrium i.e. in the range 
\[
T_{ew}\sim 100\text{ GeV}<T_{\text{sph}}\sim 10^{12}\text{ GeV} 
\]
which is of interest for baryongenesis; this is the case for all gauge
interactions. This is not always true for Yukawa interactions. The rate of a
scattering process between left and right handed fermions, Higgs bosons and
\thinspace $W$-bosons 
\[
\psi _L\phi \longrightarrow \psi _RW 
\]
is [c.f. Eq. (\ref{27}), $k_B=1$] 
\begin{equation}
\gamma \sim \alpha _2h^2\frac{\left( k_BT\right) ^5}{\left( k_BT\right) ^4}%
=\alpha _2h^2T  \label{125}
\end{equation}
The equilibrium condition is satisfied for 
\begin{eqnarray}
\gamma &>&H=1.66g_{*}^{1/2}\frac{T^2}{M_{\text{Pl}}}  \nonumber \\
T &<&\alpha _2h^2M_{\text{Pl}}\frac 1{1.66g_{*}^{1/2}}\sim h^2\frac
3{1.66g_{*}^{1/2}}10^{17}  \nonumber \\
T &<&h^210^{16}\text{ GeV}  \label{126}
\end{eqnarray}
where we have used 
\[
g^{*}=4\times 2+1+\frac 78\left( 9\times 4+2\times 4+3\times 2\right) \simeq
50 
\]
Now [$\left\langle \phi \right\rangle =v/\sqrt{2}=175$ GeV] 
\begin{eqnarray}
h_e &=&\frac{m_e}{\left\langle \phi \right\rangle }=\frac{5\times 10^{-1}%
\text{ MeV}}{\left( 250\right) \times 10^3\text{ MeV}}  \nonumber \\
&=&3\times 10^{-6}  \nonumber \\
T_e &\sim &10^{-11}\times 10^{16}\text{ GeV}\simeq 10^5\text{ GeV}  \nonumber
\\
T_u &\sim &10^7\text{ GeV}  \nonumber \\
&&\vdots  \nonumber \\
T_s &=&10^7\left( \frac{m_s}{m_u}\right) ^2\sim \left( 0.9\right)
10^{10}\simeq 10^{10}\text{ GeV}  \label{127}
\end{eqnarray}
At temperature $T\simeq 10^{10}$ GeV, which is characterestics of
leptogenesis, $e_R$, $\mu _R$, $d_R$, $s_R$ and $u_R$ are out of
equilibrium. When the Majorana right-handed $\nu $'s (these existed in early
universe) decay into leptons and Higgs scalar, 
\begin{eqnarray*}
N_R &\longrightarrow &\bar{\phi}+\ell \\
N_R &\longrightarrow &\phi +\bar{\ell}
\end{eqnarray*}
they violate lepton number. The interference between the tree-level
amplitude and the absorption part of the one-loop vertex leads to lepton
asymmetry (and baryon asymmetry $\eta _B=\frac a{a-1}\eta _L$) of the right
order of magnitude to explain the observed $\eta _B$. It has been observed
that $CP$ violation may be considerably enhanced if two heavy right handed $%
\nu $'s are nearly degenerate in mass.

As already seen $\Delta L=2$ interaction of the form 
\begin{equation}
\frac GM\left( \ell _L\ell _L\phi \phi \right) =\frac{m_\nu }{\left\langle
\phi \right\rangle ^2}\ell _L\ell _L\phi \phi  \label{129}
\end{equation}
is generated through the exchange of $N_R$ [Fig. 7].

These processes will take place with the rate 
\[
\gamma _{\Delta L=2}\left( T\right) =\frac 1{\pi ^3}\frac{T^3}{\left\langle
\phi \right\rangle ^4}\sum_{i=e,\mu ,\tau }m_{\nu _i}^2 
\]
The requirement for the harmless lepton number violation 
\[
\gamma _L<H 
\]
give 
\begin{eqnarray}
\frac 1{\pi ^3}\frac{T^3}{\left\langle \phi \right\rangle ^4}\sum_{i=e,\mu
,\tau }m_{\nu _i}^2 &<&1.66g^{*1/2}\frac{T^2}{M_{\text{Pl}}}  \nonumber
\label{130} \\
\sum_{i=e,\mu ,\tau }m_{\nu _i}^2 &<&1.66g^{*1/2}\pi ^3\frac{\left\langle
\phi \right\rangle ^4}{M_{\text{Pl}}}\frac 1T  \nonumber \\
&=&1.66g^{*1/2}\pi ^3\frac{10^9\text{ GeV}^4}{10^{19}\text{ GeV}}\frac 1T 
\nonumber \\
\sum_{i=e,\mu ,\tau }m_{\nu _i}^2 &=&<1.66g^{*1/2}\times 31\left( \frac{10^8%
\text{ GeV}}T\right) \text{eV}^2  \nonumber \\
&=&g^{*1/2}0.56\left( \frac{10^{10}\text{ GeV}}T\right) \text{eV}^2 
\nonumber \\
&=&4\left( \frac{10^{10}\text{ GeV}}T\right) \text{eV}^2  \nonumber \\
\sum_{i=e,\mu ,\tau }m_{\nu _i}^2 &\leq &\left[ 2\text{ eV}\left( \frac{T_X}{%
10^{10}\text{ GeV}}\right) ^{-1/2}\right] ^2  \label{130}
\end{eqnarray}
where 
\begin{equation}
T_X\equiv Min\left( T_{B-L}\text{, }10^{12}\text{ GeV}\right)  \label{131}
\end{equation}
$T_{B-L}$ is the temperature at which $B-L$ number production takes place.

Now $10^{12}$ is the temperature at which sphaleron transitions enter in
equilibrium. Thus 
\begin{equation}
\sum_{i=e,\mu ,\tau }m_{\nu _{i}}^{2}\leq \left[ 0.4\text{ eV}\left( \frac{%
T_{X}}{T_{\text{SPH}}}\right) ^{-1/2}\right] ^{2}  \label{132}
\end{equation}
We can reverse the argument and for $T_{B-L}\simeq 10^{16}$ GeV as in $%
SO(10) $, Eq. (\ref{130}) implies 
\begin{equation}
m_{\nu }\leq 2\text{ eV}\left( 10^{6}\right) ^{-1/2}=2\times 10^{-3}\text{ eV%
}  \label{133}
\end{equation}
which is of interest in neutrino oscillations.

\section{Thermal Leptogenesis}

One starts from a thermal distribution of heavy Majorana neutrinos which
have $CP$ violating decay modes into standard leptons: Natural candidates
are $\nu _{Ri}$, $i=1,2,3$; one in each of the three lepton families, while
the Lagrangian of electroweak interactions keep invariance under the $%
SU(2)_{L}\times U(1)_{Y}$ gauge transformations.

In this case Yukawa interactions are described by 
\begin{equation}
{\cal L}_{Y}=-\bar{\ell}_{Li}\phi h_{Lij}e_{Rj}+\bar{\ell}_{Li}\tilde{\phi}%
h_{Lij}^{*}\nu _{Rj}-\frac{1}{2}\bar{\nu}_{R}^{c}M\nu _{R}+h.c.  \label{134}
\end{equation}
the lepton number violation is induced by the third term. $M$ is a Majorana
mass matrix while $h_{L}$ are the Yukawa couplings. After spontaneous
symmetry breaking the vacuum expectation value of the\ Higgs field $%
\left\langle \phi \right\rangle =v\simeq 175$ GeV generates the Dirac mass
term $\left( m_{D}\right) _{ij}=h_{ij}v$, assumed to be small compared to $M$%
. Light neutrino mass matrix $M_{\nu }$ arises from the diagonalizating the $%
6\times 6$ neutrino mass matrix 
\begin{equation}
M_{\nu }=\left( 
\begin{array}{cc}
0 & m_{D}^{T} \\ 
m_{D} & M
\end{array}
\right)  \label{135}
\end{equation}
and takes the seesaw form 
\begin{equation}
m_{\nu }=-m_{D}^{T}M^{-1}m_{D}  \label{136}
\end{equation}
This also yields light and heavy neutrino mass eigenstates 
\begin{eqnarray}
\nu &\simeq &V_{\nu }^{T}\nu _{L}+\nu _{L}^{c}V_{\nu }^{*}  \nonumber \\
N &\simeq &\nu _{R}+\nu _{R}^{c}  \nonumber \\
m_{N_{i}} &=&M_{i}  \label{137}
\end{eqnarray}
where $V_{\nu }$ is the neutrino mixing matrix. We shall restrict our
discussion to the case of hierarchical Majorana neutrino masses, $M_{1}\ll
M_{2},M_{3}$ so that if the interactions of $N_{1}=N$ are in thermal
equilibrium when $N_{2}$ and $N_{3}$ decay, the asymmetry produced by $N_{2}$
and $N_{3}$ can be erased before $N_{1}$ decays. The asymmetry is then
generated by the out of equilibrium $CP$ violating decays of $N\rightarrow
\ell H$ versus $N\rightarrow \bar{\ell}H$ at a temperature $T\sim M\equiv
M_{1}\ll M_{2}$, $M_{3}.$

The crcuial ingredients in leptogenesis scenario is $CP$ asymmetry generated
through the interference between tree level and one-loop Majorana neutrino
decay diagrams. In the simplest extension of SM, these are shown below in
Fig. 8.

Then the $CP$ asymmetry is caused by interference between the above
diagrams: 
\begin{eqnarray}
\epsilon _{1} &=&\frac{\Gamma \left( N_{1}\rightarrow \ell _{i}H\right)
-\Gamma \left( N_{1}\rightarrow \bar{\ell}_{i}H^{*}\right) }{\Gamma \left(
N_{1}\rightarrow \ell _{i}H\right) +\Gamma \left( N_{1}\rightarrow \bar{\ell}%
_{i}H^{*}\right) }  \nonumber \\
&=&\frac{1}{8\pi }\frac{1}{\left| h_{1i}\right| ^{2}}\sum_{\ell =2,3}\func{Im%
}\left[ h_{1i}h_{1k}h_{\ell i}^{*}h_{\ell k}^{*}\right] \left[ f\left( \frac{%
M_{\ell }^{2}}{M_{1}^{2}}\right) +g\left( \frac{M_{\ell }^{2}}{M_{1}^{2}}%
\right) \right]  \label{138}
\end{eqnarray}
\begin{eqnarray}
f\left( x\right) &=&\sqrt{x}\left[ 1-\left( 1+x\right) \ln \left( \frac{1+x}{%
x}\right) \right]  \nonumber \\
&\rightarrow &\frac{\sqrt{x}}{x}\left[ -\frac{1}{2}\right] \text{, as }\frac{%
1}{x}\rightarrow 0  \label{139} \\
g\left( x\right) &=&\frac{\sqrt{x}}{1-x}\rightarrow -\frac{\sqrt{x}}{x} 
\nonumber
\end{eqnarray}
and 
\begin{equation}
I_{1l}^{ik}=\frac{1}{\left| h_{1i}\right| ^{2}}\func{Im}\left[
h_{1i}h_{1k}h_{\ell i}^{*}h_{2k}^{*}\right]  \label{140}
\end{equation}
Thus 
\begin{eqnarray}
\epsilon _{1} &=&-\frac{3}{16\pi }\left[ I_{12}^{ik}\frac{M_{1}}{M_{2}}%
+I_{13}^{ik}\frac{M_{1}}{M_{3}}\right]  \label{141} \\
I_{12}^{ik} &=&\frac{1}{\left| h_{1i}\right| ^{2}}\func{Im}\left[
h_{1i}h_{1k}h_{2i}^{*}h_{2k}^{*}\right]  \label{142}
\end{eqnarray}
The lepton asymmetry $Y_{L}$ is related to the $CP$ asymmetry through the
relation 
\begin{equation}
Y_{L}=\frac{n_{L}-\bar{n}_{L}}{s}={\cal K}\frac{\epsilon _{1}}{g^{*}}
\label{143}
\end{equation}
where $g^{*}$ is the effective number of relativistic degrees of freedom
contributing to the entropy and ${\cal K}$ is the so called dilution factor
which accounts for the wash out processes [inverse decay and lepton number
violating scattering; such processes can create a thermal population of
heavy neutrinos of high temperature $T>M$] and it can be obtained through
solving the Boltzmann equations. In the SM, $g^{*}=12\times 2+\frac{7}{8}%
\left( 18\times 4+3\times 4+3\times 2\right) =103.75.$

The produced lepton asymmetry through $Y_{L}$ is converted into a baryon
asymmetry as already seen [$a=-\frac{2}{3}$] 
\begin{eqnarray}
Y_{B} &=&\frac{a}{a-1}Y_{L}  \nonumber \\
&\simeq &0.4Y_{L}  \label{144}
\end{eqnarray}
Now 
\begin{eqnarray}
\left( m_{D}\right) _{ij} &=&h_{ij}v  \nonumber \\
\left( m_{D}m_{D}^{\dagger }\right) _{11} &=&\left( m_{D}\right) _{1i}\left(
m_{D}^{\dagger }\right) _{i1}  \nonumber \\
&=&\left( h_{1i}h_{1i}^{*}\right) v^{2}=\left| h_{1i}\right| ^{2}v^{2}
\label{145} \\
\left( m_{D}m_{D}^{\dagger }\right) _{12} &=&\left( m_{D}\right) _{1i}\left(
m_{D}^{\dagger }\right) _{i2}  \nonumber \\
&=&\left( h_{1i}h_{2i}^{*}\right) v^{2}\equiv \left( h_{1k}h_{2k}^{*}\right)
v^{2}  \nonumber
\end{eqnarray}
Thus from Eq. (\ref{142}) 
\[
I_{12}^{ik}=\frac{1}{v^{2}\left( m_{D}m_{D}^{\dagger }\right) _{11}}\func{Im}%
\left[ \left( \left( m_{D}m_{D}^{\dagger }\right) _{12}\right) ^{2}\right] 
\frac{M_{1}}{M_{2}}+\func{Im}\left[ \left( \left( m_{D}m_{D}^{\dagger
}\right) _{13}\right) ^{2}\right] \frac{M_{1}}{M_{3}} 
\]

For illustrative purposes we consider two right handed neutrinos $N_{1,2}$
and take 
\begin{equation}
m_D=\left( 
\begin{array}{ccc}
a & a^{\prime } & 0 \\ 
0 & b & b^{\prime }
\end{array}
\right)  \label{146}
\end{equation}
Then the seesaw form (\ref{136}) becomes 
\begin{eqnarray}
m_\nu &=&\left( 
\begin{array}{cc}
a & 0 \\ 
a^{\prime } & b \\ 
0 & b^{\prime }
\end{array}
\right) \left( 
\begin{array}{cc}
\frac 1{M_1} & 0 \\ 
0 & \frac 1{M_2}
\end{array}
\right) \left( 
\begin{array}{ccc}
a & a^{\prime } & 0 \\ 
0 & b & b^{\prime }
\end{array}
\right)  \nonumber \\
&=&\left( 
\begin{array}{ccc}
\frac{a^2}{M_1} & \frac{aa^{\prime }}{M_1} & 0 \\ 
\frac{aa^{\prime }}{M_1} & \frac{a^{\prime 2}}{M_1}+\frac{b^2}{M_2} & \frac{%
bb^{\prime }}{M_2} \\ 
0 & \frac{bb^{\prime }}{M_2} & \frac{b^{\prime 2}}{M_1}
\end{array}
\right)  \label{147}
\end{eqnarray}
We have to diagonalize it in order to go to mass basis: 
\begin{equation}
\left| 
\begin{array}{ccc}
\frac{a^2}{M_1}-\lambda & \frac{aa^{\prime }}{M_1} & 0 \\ 
\frac{aa^{\prime }}{M_1} & \frac{a^{\prime 2}}{M_1}+\frac{b^2}{M_2}-\lambda
& \frac{bb^{\prime }}{M_2} \\ 
0 & \frac{bb^{\prime }}{M_2} & \frac{b^{\prime 2}}{M_2}-\lambda
\end{array}
\right| =0  \label{148}
\end{equation}
Under the assumption, $\frac{b^{\prime 2}}{M_2},\frac{b^2}{M_2}\ll \frac{a^2%
}{M_1}$, this give the mass eigenvalues $0$, $\frac{b^{\prime 2}}{M_2}$,$%
\frac{a^2+a^{\prime 2}}{M_1}$. We may identify 
\begin{equation}
\Delta m_{\text{atm}}^2=m_3^2-m_1^2=\left( \frac{a^2+a^{\prime 2}}{M_1}%
\right) ^2  \label{149}
\end{equation}
\begin{equation}
\Delta m_s^2=m_2^2-m_1^2=\left( \frac{b^{\prime 2}}{M_2}\right) ^2
\label{150}
\end{equation}
so that 
\begin{eqnarray}
\frac{a^2+a^{\prime 2}}{M_1} &\simeq &\left( \Delta m_{\text{atm}}^2\right)
^{1/2}\simeq \left( 3\times 10^{-3}\text{ eV}^2\right) ^{1/2}  \nonumber \\
&\simeq &5\times 10^{-2}\text{ eV}  \label{151} \\
\frac{b^{\prime 2}}{M_2} &\simeq &\left( \Delta m_s^2\right) ^{1/2}=\left(
5\times 10^{-2}\text{ eV}^2\right) ^{1/2}  \label{152} \\
&\simeq &7\times 10^{-3}\text{ eV}  \nonumber
\end{eqnarray}
Now 
\begin{eqnarray}
\left( m_Dm_D^{\dagger }\right) _{11} &=&\left| a\right| ^2+\left| a^{\prime
}\right| ^2  \nonumber \\
\left( m_Dm_D^{\dagger }\right) _{12} &=&a^{\prime }b^{*}  \label{153}
\end{eqnarray}
Therefore [c.f. Eq. (\ref{141})] 
\begin{equation}
\epsilon _1=-\frac 3{16\pi v^2}\frac 1{\left| a\right| ^2+\left| a^{\prime
}\right| ^2}\left\{ \func{Im}\left[ \left( a^{\prime }b^{*}\right) ^2\right] 
\frac{M_1}{M_2}\right\}  \label{154}
\end{equation}
Take now $a^{\prime }=Yae^{i\delta }$ and $b^{\prime }=b$ as real. Then 
\begin{eqnarray}
\epsilon _1 &=&-\frac 3{16\pi v^2}\frac{Y^2}{1+Y^2}\frac{b^2}{M_2}\sin
2\delta  \nonumber \\
&\simeq &-\frac 3{16\pi }\frac{M_1}{v^2}\frac{Y^2}{1+Y^2}7\times 10^{-3}%
\text{ eV}\sin 2\delta  \nonumber \\
&\simeq &-1.36\times 10^{-17}\frac{M_1}{\text{GeV}}\frac{Y^2}{1+Y^2}\sin
2\delta  \label{155}
\end{eqnarray}
Taking $Y\simeq 1$, $\sin 2\delta \simeq -1$, $M_1\simeq 10^{10}$ GeV 
\begin{equation}
\epsilon _1=0.68\times 10^{-7}=6.8\times 10^{-8}  \label{156}
\end{equation}
Thus from equations (\ref{143}) and (\ref{144}) 
\begin{equation}
Y_B={\cal K}\left( 2.7\right) \times 10^{-10}  \label{157}
\end{equation}
This gives the right order magnitude as typical numbers one expects for $%
{\cal K\simeq }10^{-1}$ to $10^{-2}$ .

Acknowledgements: The author would like to acknowledge the support of King
Fahd University of Petroleum and Minerals for this work.

\section{Figure Captions}

\begin{itemize}
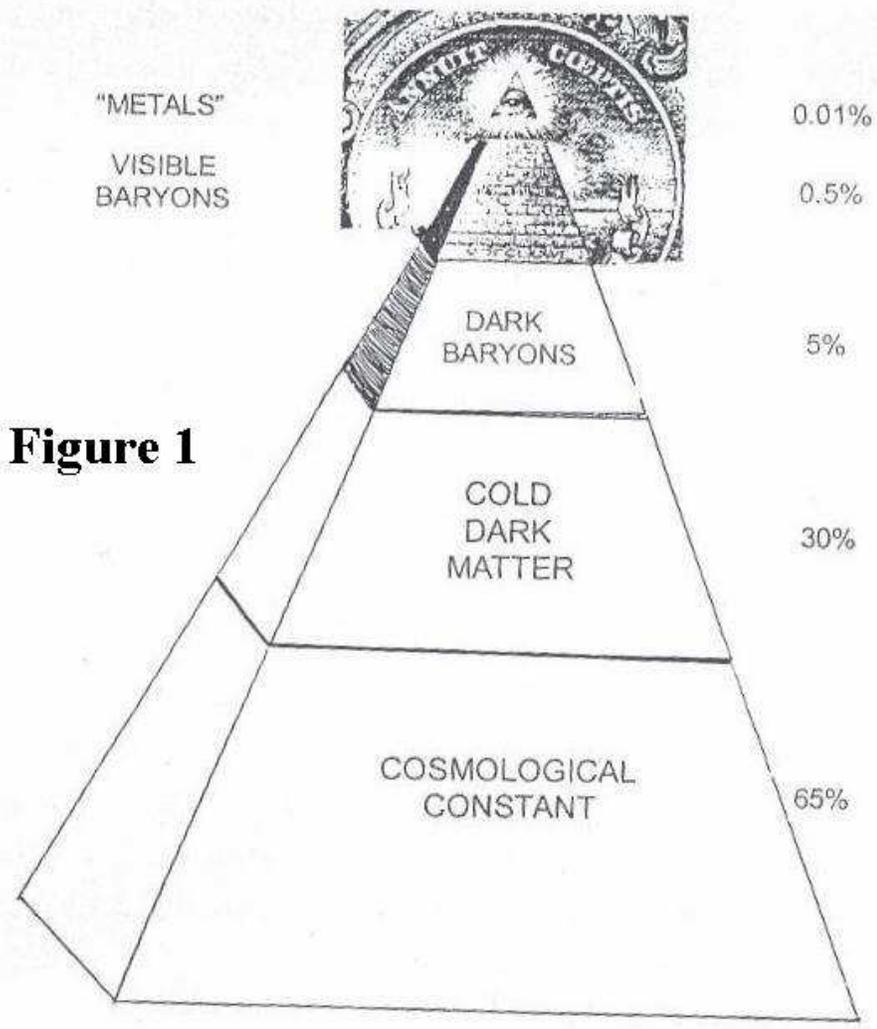
\item[Figure 1:]  Cosmic density pyramid

\item[Figure 2:]  Tree level Higgs couplings with fermions in SU(5)

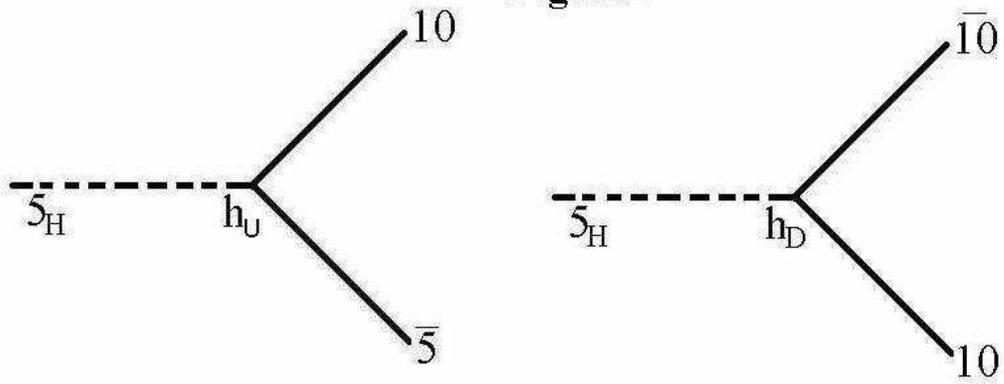
\item[Figure 3:]  Loop level Higgs and gauge particles contribution to
baryon asymmetry

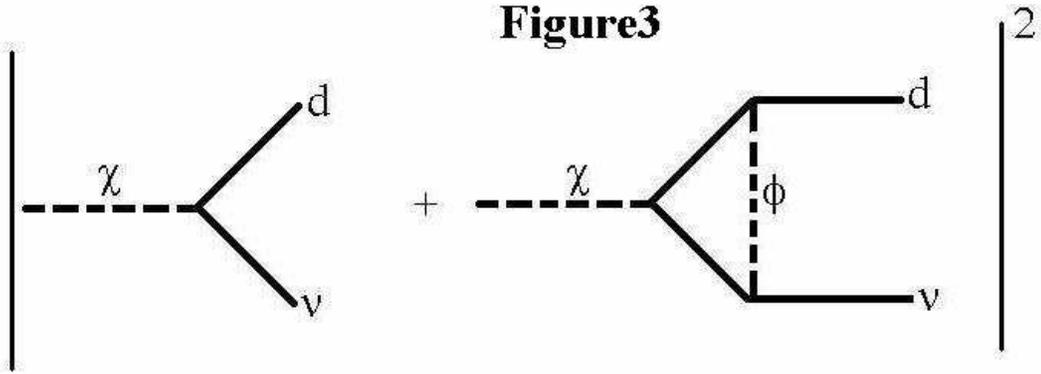
\item[Figure 4:]  Loop level contribution of $^{5}$H Higgs boson to baryon
asymmetry

\item[Figure 5:]  Transition from one vacuum to another by thermal
fluctuations

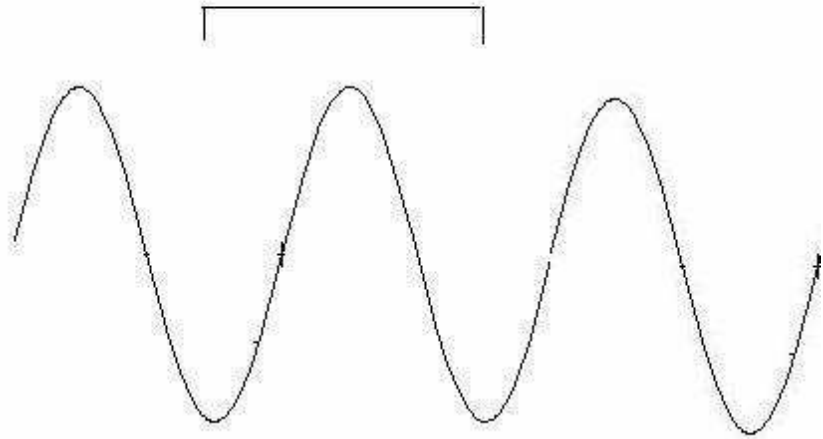
\item[Figure 6:]  Feynmann diagram for scattering process between left and
right handed fermions, Higgs bosons and $W$ boson

\item[Figure 7:]  See--saw mechenism for Majorana light neutrino masses

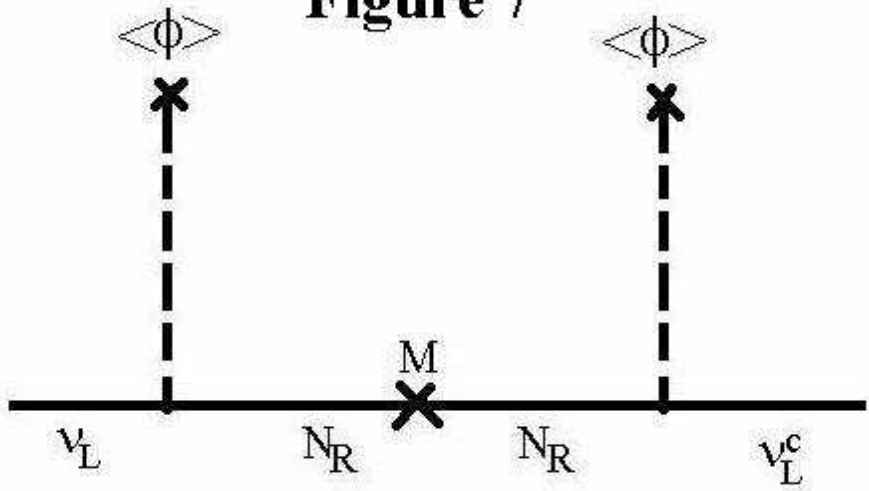
\item[Figure 8:]  Tree and one--loop level Majorana (heavy) neutrino decay
diagrams
\end{itemize}

\newpage\


\begin{thebibliography}{ }
\bibitem[ ]{gdasf}  The lectures are based on the following refrences:
\end{thebibliography}
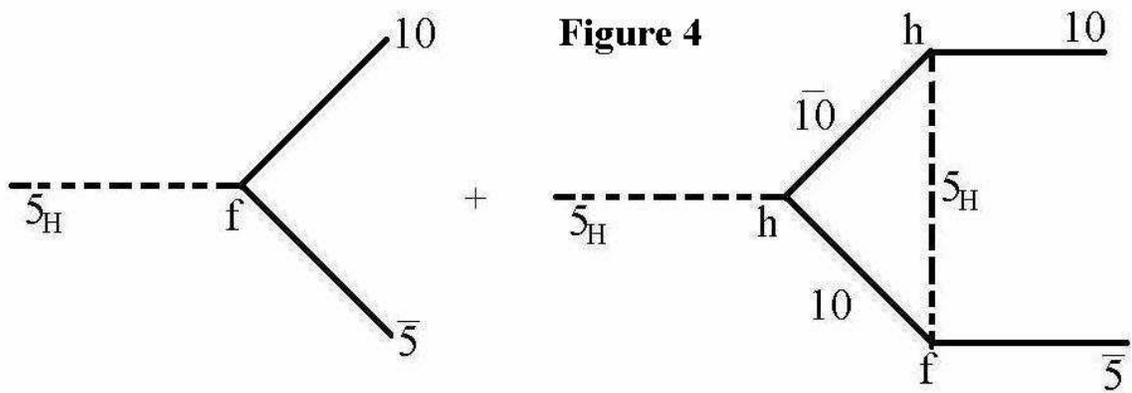
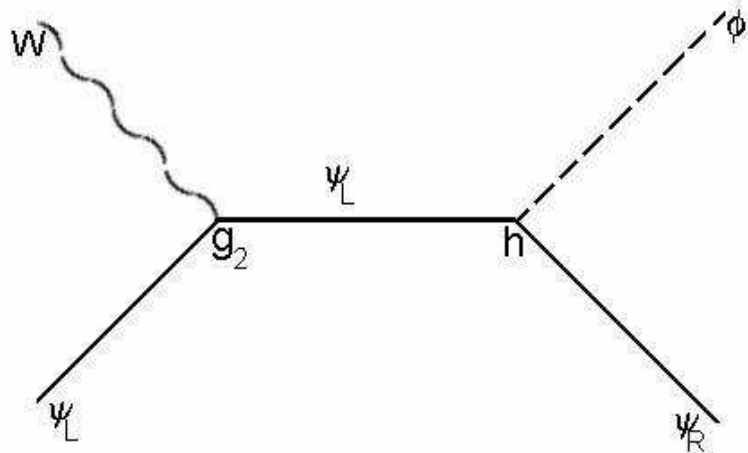
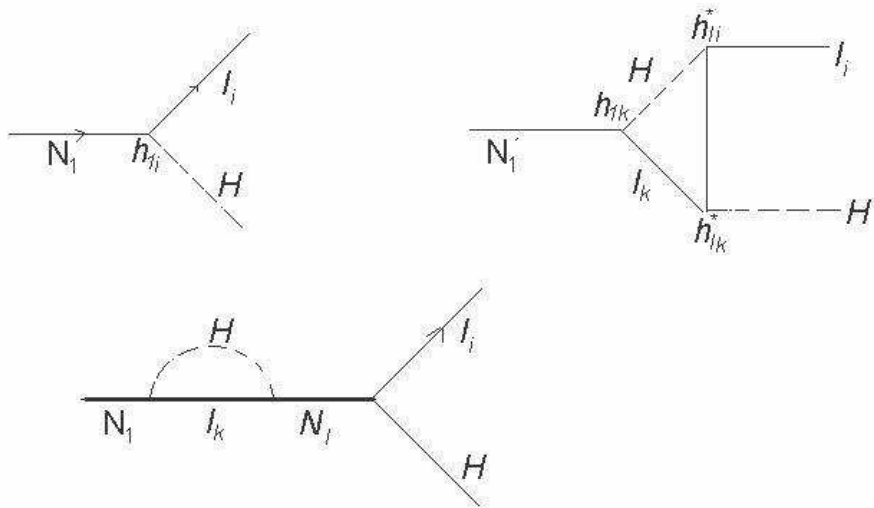
\end{document}